\documentclass[aps, floatfix,nofootinbib,amsmath,amssymb,amsfonts,notitlepage, twocolumn,superscriptaddress]{revtex4-1} 

\usepackage[T1]{fontenc}
\usepackage[latin9]{inputenc}
\usepackage{lmodern} % load a font with all the characters
\usepackage{color}
\usepackage{bbold}
\usepackage{dsfont}
\usepackage{graphicx}
\usepackage[caption=false]{subfig}
\usepackage[colorlinks]{hyperref}
\usepackage[bottom]{footmisc}
\usepackage{enumerate}
\usepackage{float}
\usepackage{mathtools}

\usepackage{empheq}
\usepackage{tikz}
\usetikzlibrary{patterns,tikzmark, matrix,decorations.pathreplacing, calc, positioning,fit}

\usepackage{hf-tikz}

\usepackage{fancyhdr}

\usepackage[normalem]{ulem}
\DeclareMathAlphabet\mbc{OMS}{cmsy}{b}{n}
\usepackage{balance}

\usepackage{diagbox}

\usepackage{comment}

\pgfmathsetmacro{\flagscalefactor}{0.3}
\pgfmathsetmacro{\flagrotationdegree}{45}

\usepackage{scalerel}

\begin{document}

\newcommand{\ks}[1]{{\textcolor{teal}{[KS: #1]}}}

\global\long\def\eqn#1{\begin{align}#1\end{align}}
\global\long\def\vec#1{\overrightarrow{#1}}
\global\long\def\ket#1{\left|#1\right\rangle }
\global\long\def\bra#1{\left\langle #1\right|}
\global\long\def\bkt#1{\left(#1\right)}
\global\long\def\sbkt#1{\left[#1\right]}
\global\long\def\cbkt#1{\left\{#1\right\}}
\global\long\def\abs#1{\left\vert#1\right\vert}
\global\long\def\cev#1{\overleftarrow{#1}}
\global\long\def\der#1#2{\frac{{d}#1}{{d}#2}}
\global\long\def\pard#1#2{\frac{{\partial}#1}{{\partial}#2}}
\global\long\def\re{\mathrm{Re}}
\global\long\def\im{\mathrm{Im}}
\global\long\def\dd{\mathrm{d}}
\global\long\def\ddd{\mathcal{D}}

\global\long\def\avg#1{\left\langle #1 \right\rangle}
\global\long\def\mr#1{\mathrm{#1}}
\global\long\def\mb#1{{\mathbf #1}}
\global\long\def\mc#1{\mathcal{#1}}
\global\long\def\tr{\mathrm{Tr}}

\global\long\def\nth{$n^{\mathrm{th}}$\,}
\global\long\def\mth{$m^{\mathrm{th}}$\,}
\global\long\def\non{\nonumber}

\newcommand{\orange}[1]{{\color{orange} {#1}}}
\newcommand{\cyan}[1]{{\color{cyan} {#1}}}
\newcommand{\teal}[1]{{\color{teal} {#1}}}
\newcommand{\blue}[1]{{\color{blue} {#1}}}
\newcommand{\yellow}[1]{{\color{yellow} {#1}}}
\newcommand{\green}[1]{{\color{green} {#1}}}
\newcommand{\red}[1]{{\color{red} {#1}}}
\newcommand{\purple}[1]{{\color{purple} {#1}}}

\global\long\def\todo#1{\cyan{{$\bigstar$ \orange{\bf\sc #1 }}$\bigstar$} }

\global\long\def\addref#1{\orange{{$\bigstar$ \cyan{\bf\sc Add reference }}$\bigstar$} }

\global\long\def\redflag#1{\Rflag{first} \red{\bf \sc #1}}

\title{Collective Quantum Beats from Distant Multilevel Emitters}

\author{Ahreum Lee}
\affiliation{Joint Quantum Institute, University of Maryland and the National Institute of Standards and Technology, College Park, Maryland 20742, USA}

\author{Hyok Sang Han}
\affiliation{Joint Quantum Institute, University of Maryland and the National Institute of Standards and Technology, College Park, Maryland 20742, USA}

\author{Fredrik K. Fatemi}
\affiliation{U.S. Army Research Laboratory, Adelphi, Maryland 20783, USA}
\affiliation{Quantum Technology Center, University of Maryland, College Park, MD 20742, USA}

\author{S. L. Rolston}
\affiliation{Joint Quantum Institute, University of Maryland and the National Institute of Standards and Technology, College Park, Maryland 20742, USA}
\affiliation{Quantum Technology Center, University of Maryland, College Park, MD 20742, USA}

\author{Kanu Sinha} 
% \email{kanu.sinha@asu.edu}
\affiliation{School of Electrical, Computer and Energy Engineering, Arizona State University, Tempe, AZ 85287-5706, USA}

\begin{abstract}
We analyze the dynamics of quantum beats in a system of two V-type three-level atoms coupled to a waveguide.  We show that quantum beats can be collectively enhanced or suppressed, akin to Dicke super- and sub-radiance, depending on the interatomic separation and the initial correlations between the atoms. In particular, the interference properties of the collective beats are determined by the distance between the atoms modulo the beat wavelength. We study  the collective atomic and field dynamics, illustrating a crossover  from a Markovian to a non-Markovian regime as the atomic separation becomes sufficiently large to bring memory effects of the electromagnetic environment into consideration. In such a non-Markovian regime, collective quantum beats can be enhanced beyond the Markovian limit as a result of retardation effects. Our results demonstrate the  rich interplay between multilevel and multiatom quantum interference effects arising in a system of distant quantum emitters.
\end{abstract}

\maketitle

\section{Introduction}

% possibly: add interesting quantum beats?
Quantum beats refer to the quantum interference effect in the  radiation emitted from different excited  levels in a multilevel atomic system \cite{ Jaynes_1980}. Similar to the well-known phenomenon of  collective atomic spontaneous emission \cite{Dicke_1954}, quantum beats can exhibit cooperative effects when considering the fluorescence from a collection of multilevel atoms as demonstrated theoretically \cite{Agarwal_1977} and experimentally \cite{Han21}. Collective effects can thus be a tool for enhancing quantum beats, relevant to improving the sensitivity of precision time-resolved spectroscopy methods \cite{Haroche1976}.

Collective atom-field interactions have been historically explored in systems where  atoms are confined within small volumes compared to the resonant wavelengths \cite{Gross76, Gross82, Skribanowitz73, Pavolini85, Devoe96}. However, waveguides allow for the realization of cooperative effects between distant emitters, which has been a subject of significant interest in recent theoretical  and experimental works \cite{Sheremet2021WaveguideQE, Pivovarov21, Asenjo17, vanLoo2013, Li16, Solano2017,Kim2018,Newman2018,Boddeti2022,Pennetta2022, Zanner22, PabloReview}. In such cases, the radiation emitted from a pair of symmetrically correlated emitters is super(sub)-radiant for an interatomic separation that is an (half-)integer multiple of the resonant transition wavelength. Thus the atomic separation $(d)$ modulo the resonant wavelength is crucial in determining the collective emission properties of a system. The interference can thus be engineered in ordered atomic arrays to exhibit strong collective phenomena creating nearly perfect mirrors \cite{ChangNJP, Corzo16, Mirhosseini19} and facilitating quantum metrology \cite{Ostermann13, Henriet19} and quantum memory \cite{Asenjo17, Manzoni_2018}.

\begin{figure}[!b]
    \centering
    \includegraphics[width = 3.4 in]{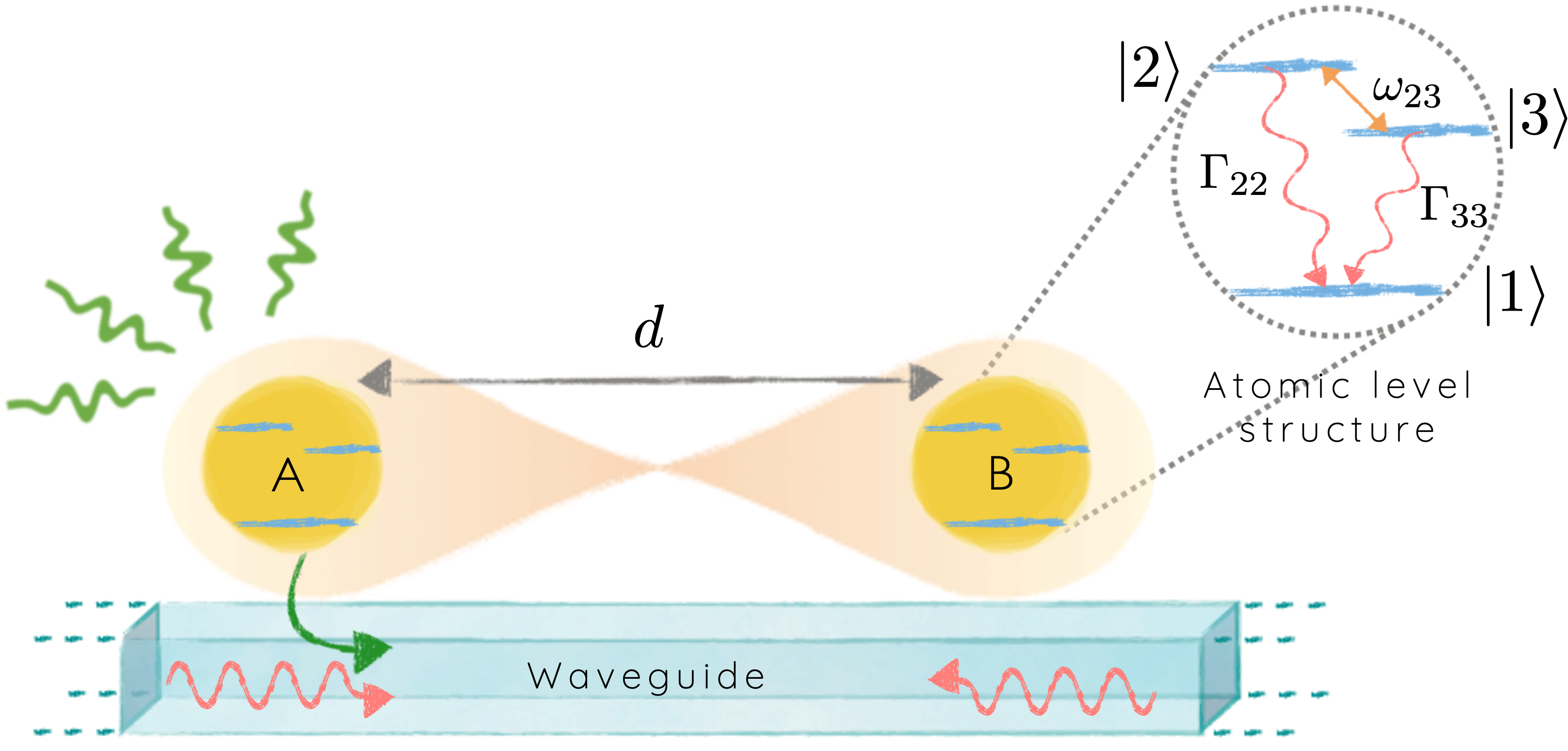}
    \caption{Schematic representation of two three-level atoms, denoted by A and B, coupled to a waveguide. We consider that each atom has a V-type level structure, with the ground state denoted by $\ket{1}$ and the two excited states denoted by $\ket{2}$ and $\ket{3}$, with decay rates $ \Gamma_{22}$ and $\Gamma_{33}$, respectively. The detuning between levels $\ket{2}$ and $\ket{3}$ is $ \omega_{23}$. We consider different regimes of the interatomic separation $ d$ such that: (1) $d \sim \lambda_{\mr{beat}} \ll L_\mr{c}$ and (2) $ \lambda_{\mr{beat}} \ll d \sim L_\mr{c}$, with $\lambda_\mr{beat } = 2\pi v/\omega_{23}$ as the beat wavelength and $L_\mr{c} = v/\Gamma_{22}$ as the coherence length, with $v$ as the speed of EM field in the waveguide.}
    \label{Fig:sch}
\end{figure}

In this work, we study the collective quantum beat dynamics of distant multilevel emitters coupled to a waveguide. In such case, the collective dynamics involves multiple transition frequencies and exhibit even richer interference behavior. We find that a larger length scale, the beat wavelength  $\lambda_\mr{beat} \equiv  2\pi v/\omega_\mr{beat}$ ($\omega_\mr{beat}$ being the beat frequency and $v$ being the speed of light in the waveguide), becomes relevant in determining the phase relations between the radiation emitted from different transitions. For a symmetrically correlated pair of distant atoms, we show that the resulting quantum beats can be enhanced(suppressed) for an interatomic separation that is an (half-)integer multiple of beat wavelength, similar to the dependence of Dicke super- and sub-radiance on the atomic separation modulo the resonant wavelength. 

Furthermore, we investigate the regime where the interatomic separation becomes comparable to the coherence length defined as $L_\mr{c} = v/\Gamma$, where $\Gamma $ is the characteristic spontaneous emission rate for individual atoms. %In such case, the timescale 
It has been shown that in such a case the system exhibits rich retardation-induced non-Markovian dynamics, with features such as collective spontaneous emission rates exceeding those of Dicke superradiance \cite{Dinc19,Rist08, Sinha20, Sinha_2020} and formation of highly delocalized atom-photon bound states \cite{Sinha_2020, Sinha2019, Calajo_2019, Trivedi21, Hughes09}. Varying the atomic separation to regimes where the retardation effects become relevant, we illustrate a crossover from a Markovian to a non-Markovian dynamics of collective quantum beats.

The rest of the paper is organized as follows. We  present the  model for the system of two V-type three level atoms coupled to a waveguide in Section~\ref{Sec:model}. Section~\ref{Sec:CollDyn} analyzes the collective quantum beat dynamics for the atomic and field degrees of freedom. In Section~\ref{Sec:Dist}, we describe the distance dependence of collective quantum beat dynamics. We discuss the conclusions and outlook of the paper in Section~\ref{Sec:Disc}.

\section{Model}
\label{Sec:model}
We consider two three-level V-type atoms  coupled through a one-dimensional waveguide, as shown in the schematic Fig.~\ref{Fig:sch}. The ground state is labeled as $\ket{1}$ and the two excited levels are $\ket{2}$ and $\ket{3}$. The frequency difference between levels $i$ and $j$ is denoted as $\omega_{ij}$. The positions of the two atoms are denoted by $x_A = - d/2$ and $x_B= d/2$.

We can write the total Hamiltonian of the system as $H=H_0+H_{AF}$, where $H_0$ is the free Hamiltonian and $H_{AF}$ represents the atom-field interaction. The free Hamiltonian is defined as:
\eqn{
  H_0 &= \sum_{m=A,B}\sum_{j=2,3} \hbar \omega_{j1} \hat{\sigma}_{m,j}^+ \hat{\sigma}_{m,j}^- \nonumber\\
  &\quad + \sum_{k} \hbar \omega_{k} \sbkt{\hat{a}_{R,k}^{\dagger} \hat{a}_{R,k}+ \hat{a}_{L,k}^{\dagger} \hat{a}_{L,k}}.}
The first term corresponds to atomic Hamiltonian where $\hat{\sigma}_{m,j}^{\pm}$ are the atomic raising and lowering operators acting on the $j^\mr{th}$ level of atom $m$. The second term is the Hamiltonian for the electromagnetic (EM) field where the creation and annihilation operators $\hat{a}_{R,k}^{\bkt{\dagger}}$ and $\hat{a}_{L,k}^{\bkt{\dagger}}$ correspond to the right and left- propagating field modes with frequency $\omega_{k}$ along the waveguide, respectively. Moving to the interaction picture with respect to $H_0$, the atom-field interaction Hamiltonian 
$\tilde H_{AF }\equiv e^{-i H_0 t/\hbar } H_{AF} \,e^{i H_0 t/\hbar }$ 
is given by:

\begin{widetext}
\eqn{
\tilde{H}_{AF} = -\sum_{m=A,B}\sum_{j=2,3}\sum_{k} \hbar g_{m,j}(\omega_k) &\cbkt{ \hat{\sigma}_{m,j}^+\sbkt{\hat{a}_{R,k}  e^{ik\cdot x_m} + \hat{a}_{L,k}  e^{-ik\cdot x_m}}e^{i(\omega_{j1}-\omega_k) t}+H.c.}
\label{eq:H_interaction}
}
\begin{comment}
\eqn{
H_{AF} = -\sum_{m=A,B}\sum_{j=2,3}\sum_{k=-\infty}^{\infty} \hbar g_{m,j}(\omega_k) &\sbkt{ \hat{\sigma}_{m,j}^+\hat{a}_{k}  e^{ik\cdot x_m} + \hat{\sigma}_{m,j}^-\hat{a}_{k}^{\dagger} e^{-ik\cdot x_m}}
\label{eq:H_interaction}
}
\eqn{
\ket{\Psi(0)}&=\frac{1}{\sqrt{2}}\bkt{\ket{21} \pm \ket{12}}
\label{eq:psi-0}
}
\eqn{
&\ket{\Psi(t)}=\sbkt{\sum_{m=A,B}\sum_{j=2,3}c_{m,j}(t)\hat{\sigma}_{m,j}^{+}+\sum_{k}c_{k}( t)\hat{a}_{k}^{\dagger} }\ket{11}\ket{\{0\}}.
\label{eq:psi-t}
}
\end{comment}
\end{widetext}
where we have employed the rotating-wave approximation. We further assume that the atom-field coupling strengths for the two atoms are equal such that $g_{A,j}(\omega_k)=g_{B,j}(\omega_k)\equiv g_{j}(\omega_k)$.
Additionally, a perfect coupling between the atoms and the waveguide is assumed, ignoring decay into other channels.

The initial state for the system is assumed to be:
\eqn{
\ket{\Psi(0)}&=\bkt{\cos{\theta}\ket{2}_A \ket{1}_B + e^{i\phi} \sin{\theta}\ket{1}_A \ket{2}_B}\otimes \ket{\{0\}}_{a,b},
\label{eq:psi-0}
}
wherein the two atoms share an excitation in level 2 and the EM field is in the vacuum state. We remark that in the absence of an initial superposition of the excited levels 2 and 3, quantum beats can be induced from the second-order vacuum coupling \cite{Hegerfeldt_1994}, as was recently demonstrated experimentally in \cite{ Han21}.
Futhermore, the above initial state readily extends to the more general initial state in the single excitation manifold where a single excitation is shared among any of the excited states and any of the two atoms.

Observing that the interaction Hamiltonian preserves the number of excitations in the atom+field system, we make the following ansatz for the time-evolved state:
\eqn{
&\ket{\Psi(t)}=\sbkt{\sum_{m=A,B}\sum_{j=2,3}c_{m,j}(t)\hat{\sigma}_{m,j}^{+}\right.\non\\
&\left.+\sum_{k}\cbkt{c_{R}(\omega_k, t)\hat{a}_{R,k}^{\dagger} + c_{L}(\omega_k, t)\hat{a}_{L,k}^{\dagger}}}\ket{1}_A \ket{1}_B\ket{\{0\}}.
\label{eq:psi-t}
}
$ c_{m,j}\bkt{t}$  denotes the excitation amplitude for the $m^\mr{th}$ atom in the $j^\mr{th}$ level and $c_{R(L)}(\omega_k,t) $ stands for the excitation amplitude for the right(left) propagating field mode of frequency $\omega_k$.

\section{Collective Quantum Beat Dynamics}
\label{Sec:CollDyn}

\subsection{Equations of Motion}
\label{Sec:EOM}

From the interaction Hamiltonian and the single-excitation ansatz for the total system state (Eqs.~\eqref{eq:H_interaction}  and \eqref{eq:psi-t}), we obtain the equations of motion for the atomic and field excitation amplitudes as follows:
\eqn{
	\label{eq:schrodinger1}
\partial_t c_{m,j}(t) =&i\sum_{k} g_j(\omega_k) e^{i(\omega_{j1}-\omega_{k})t}\non\\
	&\sbkt{c_R\bkt{\omega_k,t}e^{ik\cdot x_m} +c_L\bkt{\omega_k,t}e^{-ik\cdot x_m} },
	}
	\eqn{
	\label{eq:EOMa}
	 \partial_t c_R\bkt{\omega_k,t} &=\non\\
	 &i\sum_{m=A,B}\sum_{j=2,3}g_j(\omega_k)e^{-i(\omega_{j1}-\omega_{k})t}   c_{m,j}(t) e^{-ik\cdot x_m},\\
	 \label{eq:EOMb}
	 \partial_t c_L\bkt{\omega_k,t} &=\non\\
	 &i\sum_{m=A,B}\sum_{j=2,3}g_j(\omega_k)e^{-i(\omega_{j1}-\omega_{k})t}   c_{m,j}(t) e^{ik\cdot x_m}.
	}

One can solve for the atomic dynamics by tracing out the field modes to obtain:

\eqn{
	\partial_t c_{m,j}(t) 	= &- \sum_{n=A,B}\sum_{l=2,3} \frac{ \Gamma_{jl}}{2} e^{i\omega_{jl}t} e^{i\omega_{l1}\frac{\abs{x_m-x_n}}{v}}\non\\
	&c_{n,l}\bkt{t-\frac{\abs{x_m-x_n}}{v}}\Theta\bkt{t-\frac{\abs{x_m-x_n}}{v}},
\label{eq:de}
}
where  we have assumed a  flat spectral density  of the EM field such that $g_{j}\bkt{\omega_k}\approx
g_{j}\bkt{\omega_{j1}}\equiv g_j$. The generalized decay rate $\Gamma_{jl}$ is defined as
\eqn{
\Gamma_{jl}=\frac{d_{j1}d_{l1}\omega_{jl}^3}{3\pi\epsilon_0 h v^3},
}
assuming the transition dipole moments are parallel to each other. 

One can identify the various processes that contribute to the total collective quantum beat dynamics from Eq.~\eqref{eq:de} as follows: 
\begin{itemize}
    \item{Individual atomic spontaneous emission:  corresponding to the terms involving the same atom and same excited level ($n = m$, $j = l$).}
    \item{Individual atomic quantum beats:  corresponding to the terms involving the same atom and different excited levels ($n = m$, $j \neq l$).}
    \item{Collective spontaneous emission: corresponding to the terms with different atoms and same  excited levels ($n\neq m$, $j = l$)}
    \item{Collective emission of quantum beats: represented by the interference terms between  different atoms and different excited levels ($n\neq m$, $j \neq l$).}
\end{itemize}

\begin{figure}[t]
    \centering
    \includegraphics[width = 3.3 in]{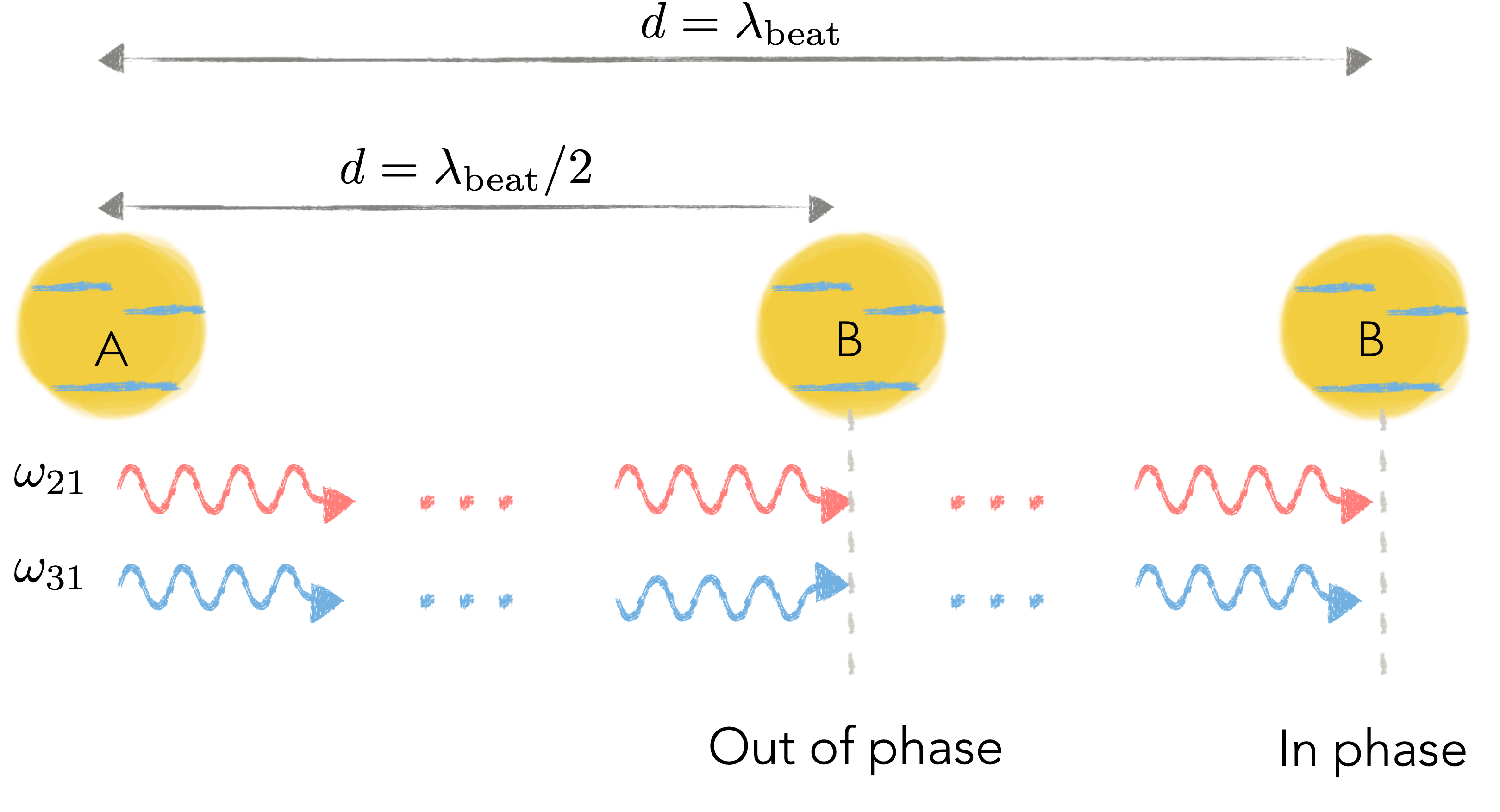}
    \caption{Schematic representation of the interference between radiation emitted from different atomic transitions for propagation distances of $d = \lambda_\mr{beat}/2$ and $ d = \lambda_\mr{beat}$. The two field modes at different frequencies are in-phase right after being emitted and gradually become out-of-phase as they travel through the waveguide. For a propagation distance of half the beat wavelength, the two modes are exactly out-of-phase with each other; for a propagation distance equal to the beat wavelength,  they become in-phase again. Thus, the interatomic distance modulo the beat wavelength determines  the interference properties of the  radiation emitted from the two transitions.}
    \label{Fig:beatphase}
\end{figure}

The collective atomic and field dynamics is obtained as a combination of the above four processes, exhibiting a rich interplay between different length scales. For example, for a symmetric initial state: 1) When $d$ is an (half-)integer multiple of the transition wavelength $\lambda_{j1}$, the photons emitted by the two distant atoms from the respective transitions $(\ket{j}\leftrightarrow\ket{1})$  are in(out-of)-phase. 
2) When $d$ is an (half-)integer multiple of the beat wavelength $\lambda_\mr{beat}$, the two photons of wavelengths $\lambda_{21}$ and $\lambda_{31}$ become in(out-of)-phase at the position of the other atom, as illustrated in Fig.~\ref{Fig:beatphase}.

Furthermore, non-Markovian retardation effects become prominent as the atomic separation becomes comparable to the coherence length of the photons emitted from the atoms. For example,  in a regime where $d\gtrsim L_c$, the time scale over which the two atoms interact via the EM field ($\sim d/v$) becomes comparable to the system relaxation time scale ($\sim 1/\Gamma_{jj}$). Thus, it is pertinent to include the retardation effects in the EM field mediating the interaction between the two atoms, making the system non-Markovian \cite{Sinha_2020}.

\subsection{Atomic Dynamics}

An arbitrary initial state  with a shared single excitation between the two atoms in level $\ket{2}$  (Eq.~\eqref{eq:psi-0}) can be always decomposed into symmetric($\ket{\psi_+}$) and anti-symmetric($\ket{\psi_-}$) states:
\eqn{
\label{eq:psi-0-pm}
\ket{\psi_\pm} = \frac{1}{\sqrt{2}}\bkt{\ket{2}_A \ket{1}_B \pm \ket{1}_A \ket{2}_B}.
}
Thus, we will limit our investigation to the initial states $\ket{\psi_\pm}$. For completeness,  the description of a general initial state case is given in Appendix~\ref{App:dyn}.

The time-evolved excitation amplitudes for the two atoms follow the symmetry of the initial state, such that:
\eqn{
  c_{A,j}^{( \pm)}(t)=\pm c_{B,j}^{( \pm)}(t),
}
where the superscripts  $+ (-)$ correspond to the (anti-)symmetric initial states $\ket{\psi_\pm}$. Importantly, we note that the symmetry of the  atomic state with respect to both the $\ket{3}\leftrightarrow\ket{1}$ and the $\ket{2}\leftrightarrow\ket{1}$ transitions is the same as the initial symmetry for the $\ket{2}\leftrightarrow\ket{1}$ transition throughout the dynamics.

To simplify the notation, we will drop the atomic  labels $m$, i.e.,  $c_{A,j}^{(\pm)}(t) \equiv c_j^{(\pm)}(t)$, and focus on the evolution of atom A. From solving the coupled atomic dynamical equations of motion in  Eq.~\eqref{eq:de} by taking Laplace transform, one obtains the atomic dynamics as the sum of various modes (see Appendix~\ref{App:dyn} for details):
\eqn{
\label{eq:c2t}
c_{2}^{(\pm)}\bkt{t} &= \sum_{n = -\infty}^{\infty} \alpha_n^{\bkt{\pm}} e^{s_n^{\bkt{\pm}} t},\\
\label{eq:c3t}
c_{3}^{(\pm)}\bkt{t} &= \sum_{n = -\infty}^{\infty} \beta_n^{\bkt{\pm}} e^{\bkt{s_n^{\bkt{\pm}}-i\omega_{23}} t}.
}
The coefficients $s_n^{\bkt{\pm}}$ and $s_n^{\bkt{\pm}} - i \omega_{23}$ denote the characteristic complex eigenfrequencies of the amplitude dynamics for the levels 2 and 3, respectively, and are defined as the poles of the propagator $G^{(\pm)}(s)$:
\begin{widetext}
\eqn{
    G^{(\pm)}(s)&\equiv\sbkt{\bkt{s-i\omega_{23}+\frac{\Gamma_{33}}{2}\pm\frac{\Gamma_{33}}{2}e^{i\phi_2}e^{-\frac{d}{v}s}}\bkt{s+\frac{\Gamma_{22}}{2}\pm\frac{\Gamma_{22}}{2}e^{i\phi_2}e^{-\frac{d}{v}s}}-\frac{\Gamma_{23}\Gamma_{32}}{4}\bkt{1\pm e^{i\phi_2}e^{-\frac{d}{v}s}}^2}^{-1}.
    \label{eq:denominator}
}
\end{widetext}
Here, $\phi_2=\omega_{21} d / v$ is the propagation phases for  the resonant transition frequency $\omega_{21}$.  In general, the propagator above has an infinite number of poles, and it is difficult to express the corresponding eigenfrequencies  in a closed-form analytical solution. We therefore obtain $s_n^{(\pm)}$ numerically for a  finite number of poles of the propagator $G^{(\pm)} \bkt{s_n^{(\pm)}}$ with $n\in \cbkt{-N,\dots , N}$.

The corresponding coefficients  $\alpha_n^{(\pm)}$ and $\beta_n^{(\pm)}$ for the $n^\mr{th}$ eigenfrequency are:
\eqn{
    \alpha_n^{(\pm)}&=\frac{1}{\sqrt{2}}\lim_{s\to s_n}\frac{s-i\omega_{23}+\frac{\Gamma_{33}}{2}\pm\frac{\Gamma_{33}}{2}e^{i\phi_2}e^{-\frac{d}{v}s}}{\partial_s\sbkt{\bkt{ {G^{(\pm)}(s)}}^{-1}}},\\
    \beta_n^{(\pm)}&=-\frac{1}{\sqrt{2}}\frac{\Gamma_{32}}{2}\lim_{s\to s_n}\frac{\bkt{1 \pm e^{i\phi_2}e^{-\frac{d}{v}s}}}{\partial_s\sbkt{\bkt{ {G^{(\pm)}(s)}}^{-1}}}.
\label{eq:alpha}
}

In the limit where atoms are co-located, the eigenfrequencies given by the poles of the propagator   (Eq.~\eqref{eq:denominator}) can be solved analytically, and the  atomic dynamics corresponds to simple collective quantum beat dynamics without delay. For the symmetric initial state, this yields two solutions: $s^{(+)}=\cbkt{-\frac{\Gamma_{22}+\Gamma_{33}}{2}+i\frac{\omega_{23}+\delta}{2}, -\frac{\Gamma_{22}+\Gamma_{33}}{2}+i\frac{\omega_{23}-\delta}{2}}$
%the complex beating frequency as
where $\delta=\sbkt{{\omega_{23}}^2-(\Gamma_{22}+\Gamma_{33})^2-2i\omega_{23}(\Gamma_{22}-\Gamma_{33})}^{\frac{1}{2}}$. In the regime where the excited levels are well-separated $\bkt{\frac{\Gamma_{ij}}{\omega_{23}}\ll1}$, the atomic dynamics can be simplified as follows:
\eqn{
\label{eq:atomic_amplitude_zero_distance_2}
 c_2^{(+)}\bkt{t} &= \frac{1}{\sqrt{2}}\sbkt{e^{-\Gamma_{22}t}-\bkt{\frac{\Gamma_{32}\Gamma_{23}}{{\omega_{23}}^2}} e^{-\Gamma_{33}t}e^{i\omega_{23}t}},\\
\label{eq:atomic_amplitude_zero_distance_3}
 c_3^{(+)}\bkt{t} &= \frac{i\Gamma_{32}}{\sqrt{2}\omega_{23}}\sbkt{e^{-\Gamma_{33}t}-e^{-\Gamma_{22}t}e^{-i\omega_{23}t}}.
}
The dynamics of the amplitude of level 2 is dominated by the collective decay at a rate $\Gamma_{22}$, overlaid with a beating term with an amplitude $\Gamma_{32}\Gamma_{23}/\omega_{23}^2\ll1$. 
The initial population in level 3 being zero, the excitations in level 3 arise exclusively from a second-order vacuum-induced coupling between level 2 and level 3. Thus, the decay and the beat terms in $c_3(t)$ have the same amplitude.

For the anti-symmetric initial state in the zero-distance case we obtain the complex eigenfrequencies as $s^{(-)}=\cbkt{0,\,i\omega_{23}}$, without any real component or decay. Thus, the system remains in the subradiant state with no evolution of the corresponding atomic coefficients: $c_2^{(-)}(t)=\frac{1}{\sqrt{2}}, c_3^{(-)}(t)=0$.

\subsection{Field Dynamics}

While the atomic dynamics provides physical intuition, it cannot be measured directly in  experiments. Instead, one measures the intensity of the light emitted from the system, which carries indirect information about the atomic dynamics.  The intensity emitted by the atomic system is given by $ I(x,t)=\frac{\epsilon_0 c}{2}\bra{\psi(t)}\hat{E}^{\dagger}(x,t)\hat{E}(x,t)\ket{\psi(t)}$, with  the electric field operator defined as 
$ \hat{E}(x,t)=\int_{0}^{\infty} \dd{k} \mc{E}_k\sbkt{ \hat{a}_{R,k} e^{ikx} + \hat{a}_{L,k}e^{-ikx}}e^{-i\omega_k t}$.  This can be calculated explicitly as follows (see Appendix~\ref{App:Int} for details):

\begin{widetext}

\eqn{\label{Eq:int}
I\bkt{x,t}/I_0 = &\left| \sum_{j = 2,3}\sum_{m = A, B} g_j \left\{\underbrace{ c_{m,j}\bkt{t-\frac{x-x_m}{v}} e^{-i\omega_{j1}\bkt{t-\frac{x - x_m}{v} }}\sbkt{ \Theta \bkt{t-\frac{x-x_m}{v}} -\Theta \bkt{-\frac{x-x_m}{v}} }}_\text{Right light cone for atom $m$ at frequency $\omega_{j1}$ }\right.\right.\non\\
&\qquad\qquad\quad\left.\left. +\underbrace{c_{m,j}\bkt{t+\frac{x-x_m}{v}} e^{-i \omega_{j1} \bkt{t+\frac{x-x_m}{v} }}\sbkt{ \Theta \bkt{t+\frac{x-x_m}{v}} -\Theta \bkt{\frac{x-x_m}{v}} }}_\text{Left light cone for atom $m$ at frequency $\omega_{j1}$} \right\}\right|^2.}
\end{widetext}
The first and second terms in the above expression correspond to the light cone emitted by atom $m$ at frequency $\omega_{j1}$ towards right and left, respectively. Before the two light cones meet, the atoms are causally disconnected and emit independently. As each atom `sees' the other atom, the interference between the light cones at the two transition frequencies emitted by the two atoms leads to collective quantum beat dynamics.

The intensity measured outside the system at $x\rightarrow x_B^+$ expressed in terms of the various system eigenfrequencies reads as:
\eqn{
 I\bkt{t}/I_0 &= \left| \sum_n \bkt{g_2\,\alpha_n^{(\pm)}+g_3\,\beta_n^{(\pm)}}\right.\non\\
 &\qquad\left.\bkt{\Theta(t)\pm e^{-s_n^{(\pm)}d/v}\Theta\bkt{t-d/v}} e^{s_n^{(\pm)}t}\right|^2.
}
From the above expression we note that  quantum beats result from the interference of the  modes with different frequencies, such that $\im s_n^{(\pm)}\neq \im s_m^{(\pm)}$. In particular, collective quantum beats originate from the interference between the fields emitted by the two atoms at different frequencies for $ t>d/v$. The collective beat amplitude has a distance dependence as can be seen from the prefactor $e^{-s_n^{(\pm)}d/v}$, which corresponds to the difference in phase and amplitude for various field modes as they propagate between the two atoms.

In the limit of two coincident atoms $(d\rightarrow0)$, the intensity measured at $x\rightarrow x_B^+$ is
\eqn{
 I\bkt{t}/I_0 = &\left|  g_2\,c_{A2}\bkt{t}+g_3\,c_{A3}\bkt{t} e^{i\omega_{23}t}\right.\non\\
 &\quad\left. + g_2\,c_{B2}\bkt{t} + g_3\,c_{B3}\bkt{t} e^{i \omega_{23} t} \right|^2\Theta \bkt{t}.
 \label{eq:intensity_zero_distance}
}
For the anti-symmetric initial state where $c_{A2}(t)=-c_{B2}(t)$ and $c_{A3}(t)=-c_{B3}(t)$, the total emitted intensity vanishes, as expected for a Dicke subradiant state.

For a  symmetric initial state where $c_{A2}(t)=c_{B2}(t)$ and $c_{A3}(t)=c_{B3}(t)$, the emitted intensity is four times that of a single three-level atom, corresponding to standard Dicke superradiance. Plugging in Eqs.~\ref{eq:atomic_amplitude_zero_distance_2} and \ref{eq:atomic_amplitude_zero_distance_3} into Eq.~\ref{eq:intensity_zero_distance}, one gets
\eqn{
 I(t)/I_0' = &\Gamma_{22}e^{-2\Gamma_{22}t} + \Gamma_{33}\frac{\Gamma_{23}\Gamma_{32}}{{\omega_{23}}^2}e^{-2\Gamma_{33}t}\non\\ 
 &- 2\frac{\Gamma_{23}\Gamma_{32}}{\omega_{23}}\sin(\omega_{23}t)e^{-(\Gamma_{22}+\Gamma_{33})t},
 \label{eq:intensity_zero_distance_2}
}
using the relation ${g_j}^2\propto \Gamma_{jj}$. The first two terms correspond to spontaneous emission from levels 2 and 3, and the third terms represents quantum beats. The above expression is in  agreement with the collective quantum beat dynamics from  co-located atoms as previously obtained in \cite{Han21}.

\section{Distance Dependence of Collective Quantum Beat Dynamics}
\label{Sec:Dist}

\begin{center}
\begin{table}[b]
\begin{tabular}{ c c }
 \hline
 \hline
 %Decay rate of level 2 ($\Gamma_{22}$) & 1 \\
 Decay rate of level 3 ($\Gamma_{33}/\Gamma_{22}$) & 1 \\
 Energy separation of level 2 and 3 ($\omega_{23}/\Gamma_{22}$) & 50 \\%$10 \sim 100$
 Resonant frequency of level 2 ($\omega_{21}/\Gamma_{22}$) & $10^4$ \\
 %Field velocity in waveguide ($v$) & 1 \\
 Coherence length ($L_\mr{c} \cdot\Gamma_{22}/v$) & 1 \\
 Beat wavelength ($\lambda_{\mr{beat}}\cdot\Gamma_{22}/v$) & $ 4\pi\times 10^{-2}$ \\
 Transition wavelength  ($\lambda_{21}\cdot\Gamma_{22}/v$) & $2\pi\times 10^{-4}$ \\
 \hline
 \hline
\end{tabular}
\caption{Summary of parameters used in  calculations, based on typical values in a superconducting circuit setup. The frequencies are in the units of $\Gamma_{22}$, and the lengths are in the units of $v/\Gamma_{22}$.
}
\label{table-parameters}
\end{table}
\end{center}

We now present the collective quantum beat dynamics for a specific implementation of the model in a superconducting circuit setup as an example \cite{Srinivasan11}, with parameters described in Table~\ref{table-parameters}. In particular we discuss the dependence of the collective system dynamics on atomic separation across two different regimes wherein (1) $d\ll L_c$ (Markovian regime) and  (2) $d\gtrsim L_c$ (non-Markovian regime). 

In each of these regimes, we analyze the system dynamics for the symmetric and anti-symmetric initial states of the two atoms (Eq.~\eqref{eq:psi-0-pm}), considering interatomic separations  of integer and half-integer multiples of the beat wavelength. For simplicity we assume that $d$ is an integer multiple of $\lambda_{21}$. While the initial state determines the total collective spontaneous emission,  the interatomic distance modulo the beat wavelength determines  the interference properties of the  collective quantum beats, as discussed in Section~\ref{Sec:EOM}.

\subsection{Markovian Regime}
\label{Sec:dist_mark}
\begin{figure}[]
    \centering
    \includegraphics[width = 3.4 in]{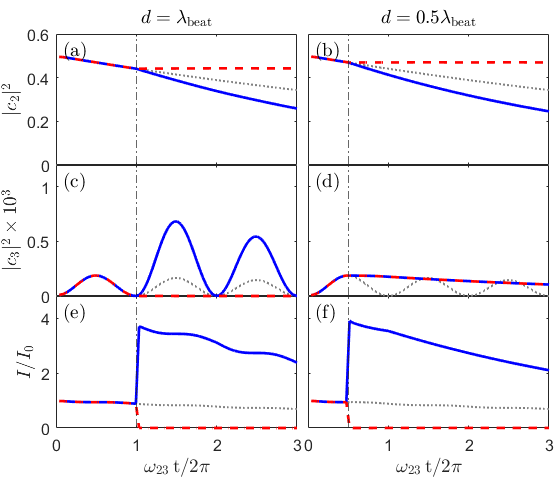}
    \caption{Atom-field dynamics in the Markovian regime: (a-b) level 2 dynamics, (c-d) level 3 dynamics, and (e-f) field dynamics measured at $x\rightarrow x_B^+$ for interatomic separations (a,c,e) $d=\lambda_\mr{beat}$ and (b,d,f) $d=\frac{1}{2}\lambda_\mr{beat}$. The solid blue curves are for symmetric initial state, and the dashed red curves for anti-symmetric initial state. The vertical dash-dotted lines indicate the times when the field emitted from one atom reaches the other atom. For comparison the single atom dynamics is drawn with the dotted gray lines. The level 3 population is scaled by a factor of $10^3$ for clarity of illustration.}
    \label{Fig:markovian_dynamics}
\end{figure}

We study the dynamics of  atomic excitation probabilities  and the field intensity for interatomic separations of $d= \cbkt{\lambda_\mr{beat}, \lambda_\mr{beat}/2}\ll L_c$. We numerically calculate the poles of Eq.~\ref{eq:denominator} for $\abs{\re\sbkt{s_n^{(j,\pm)}}} < 200\,\Gamma_{22}$ and $\abs{\im\sbkt{s_n^{(j,\pm)}}} < 200\,\omega_{23}$. Including sufficient high frequency modes allows one to correctly capture the dynamics of the system for the time scales of interest (see  Appendix~\ref{App:poles} for details).

Fig.~\ref{Fig:markovian_dynamics} depicts the atomic and field dynamics for initial symmetric and anti-symmetric states of the two atoms.
We note that the onset of collective dynamics happens at $d=\lambda_\mr{beat}$ or $d=\frac{1}{2}\lambda_\mr{beat}$ depending on the interatomic separation as indicated by the vertical dashed-dotted lines.
Fig.~\ref{Fig:markovian_dynamics} (a) and (b) illustrate the level 2 population dynamics which exhibits a super-(sub-)radiant decay for (anti-)symmetric initial states. It can be seen from Eq.~\ref{eq:atomic_amplitude_zero_distance_2} that the amplitude of the beat term is smaller compared to that of the individual decay in level 2 dynamics by a factor of $\Gamma_{23}\Gamma_{32}/{\omega_{23}}^2\ll 1$. Thus, we do not see any visible beats in the level 2 population curves.

More interestingly, Fig.~\ref{Fig:markovian_dynamics} (c) and (d) illustrate the collective quantum beat effect as seen in the level 3 population dynamics. 
We note that for an interatomic separation of  $d=\lambda_\mr{beat}$ there is a collective enhancement of the quantum beats for the symmetric initial state,  and suppression for the antisymmetric initial state as denoted in Fig.~\ref{Fig:markovian_dynamics} (c). For this separation, the phase of the field modes  mediating the interaction between the atoms is an even integer multiple of $2 \pi$ such that  $\omega_{21}d/v =2 n \pi$, $\omega_{31}d/v = 2 m \pi$ ($\cbkt{n,m}\in \mathbb N$). Furthermore, the total atomic state is (anti-)symmetric with respect to the $\ket{3}\leftrightarrow\ket{1}$ transition for an initial (anti-)symmetric state with respect to $\ket{2}\leftrightarrow\ket{1}$ transition. Thus, we observe an enhancement or suppression of the quantum beats for an initial symmetric or  anti-symmetric state, respectively. More specifically, it can be seen that the amplitude of the first peak of the collective `superradiant' quantum beats (solid blue) is enhanced roughly by a factor of $\approx 4.1$ in comparison with the independent emission case (dotted gray).
In contrast, it can be seen in Fig.~\ref{Fig:markovian_dynamics} (d) that for a separation of $d=\lambda_\mr{beat}/2$,  the resulting beats  are suppressed as a result of the destructive interference between the fields emitted from the two atoms at $\omega_{21}$ and $\omega_{31}$, as illustrated in  Fig.~\ref{Fig:beatphase}.

In Fig.~\ref{Fig:markovian_dynamics} (e) and (f), the intensity of the light measured outside the system is depicted. The radiated intensity, as given by Eq.~\eqref{Eq:int},  is governed by the interference between the atomic excitation amplitudes. For the (anti-)symmetric initial state, the overall emission is superradiant(subradiant). For  the case of superradiant emission, the size of the beat is enhanced(suppressed) for an atomic separation of  $d=\lambda_\mr{beat}$ $\bkt{d=\frac{1}{2}\lambda_\mr{beat}}$, in agreement with the collective atomic dynamics.

\subsection{Non-Markovian Regime}
\label{Sec:dist_nm}

\begin{figure}[]
    \centering
    \includegraphics[width = 3.4 in]{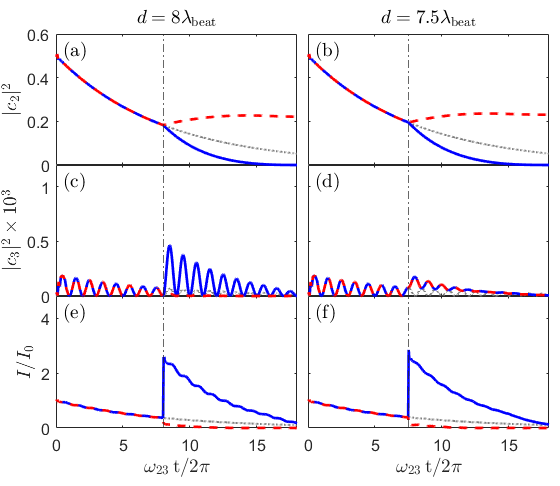}
    \caption{ Non-Markovian regime: (a-b) level 2 dynamics, (c-d) level 3 dynamics, and (e-f) field dynamics measured at $x\rightarrow x_B^+$ for interatomic separations (a,c,e) $d=8\lambda_\mr{beat}$ and (b,d,f) $d=7.5\lambda_\mr{beat}$. The solid blue curves are for symmetric initial state , and the dashed red curves for anti-symmetric initial state. The vertical dashed-dotted lines indicate the times when the field emitted from one atom reaches the other atom. For comparison the single atom dynamics is drawn with the dotted gray lines. The level 3 population is scaled by a factor of $10^3$.
    }
    \label{fig:nonmarkovian_dynamics}
\end{figure}

We now consider the case wherein the atomic separations are comparable to the coherence length of the emitted photons, with  $d=7.5 \lambda_\mr{beat}\approx 0.94 L_c$ and $d=8\lambda_\mr{beat}\approx L_c$. We note that for such large separations, the retardation  effects of the waveguide field become relevant, rendering the system dynamics non-Markovian. We calculate the dynamics numerically by obtaining the characteristic system frequencies as the poles of the propagator Eq.~\ref{eq:denominator} within  $\abs{\re\sbkt{s_n^{(j,\pm)}}} < 10\,\Gamma_{22}$ and $\abs{\im\sbkt{s_n^{(j,\pm)}}} < 10\,\omega_{23}$ (see  Appendix~\ref{App:poles} for details).

The atomic and field dynamics for this regime is shown in Fig.~\ref{fig:nonmarkovian_dynamics} for the initial symmetric and antisymmetric states.
The level 2 dynamics for a symmetric initial state as denoted by the solid blue curves in Fig.~\ref{fig:nonmarkovian_dynamics} (a) and (b) exhibits  collective emission rates faster than standard Dicke superradiance (`superduperradiance'), similar to the non-Markovian collective dynamics for a system of two two-level atoms \cite{Sinha_2020, Dinc18, Dinc19}. For an anti-symmetric initial state (dashed red curves), one can see the formation of delocalized atom-photon bound states in continuum (BIC)  \cite{Calajo_2019, Sinha2019}.

The effects of retardation on  collective quantum beats are illustrated in the population dynamics of level 3 in Fig.~\ref{fig:nonmarkovian_dynamics} (c) and (d). For an interatomic separation $d=8\lambda_\mr{beat}$ as seen in Fig.~\ref{fig:nonmarkovian_dynamics} (c), we observe an enhancement of the quantum beats for a symmetric initial state and moderate suppression of beats for the antisymmetric initial state. Furthermore, comparing the first peak of the collective quantum beats (solid blue) with that of the independent decay (dotted gray) shows an enhancement beyond the Markovian case by roughly a factor of $\sim 6.8$. 
For a separation of $d=7.5\lambda_\mr{beat}$ as illustrated in Fig.~\ref{fig:nonmarkovian_dynamics} (d), the population dynamics of level 3 for both the initial symmetric and antisymmetric cases exhibits beating in addition to an exponential decay. 

The light intensity measured outside the system is depicted in Fig.~\ref{fig:nonmarkovian_dynamics} (e) and (f). The symmetric initial state exhibits an overall exponential decay faster than Dicke superradiance, with an overlaid beating that is enhanced(suppressed) for a separation $d=8\lambda_\mr{beat}$ $\bkt{d=7.5\lambda_\mr{beat}}$. The antisymmetric initial state shows a suppressed total emission outside the system, indicating that most of the light is trapped in between the atoms forming a  delocalized atom-photon bound state.  However, there is a finite emission into the field modes from the otherwise subradiant state, indicating existence of additional modes in a non-Markovian regime that provide channels for the atomic excitations to decay away. Such modes have been previously investigated in the context of steady-state atomic spectrum emitted from two distant two-level atoms \cite{Sinha20}.

\section{Discussion}
\label{Sec:Disc}

In this work we have demonstrated the distance-dependent dynamics of collective quantum beats, resulting from the interference between the radiation emitted from a collection of multilevel atoms coupled to a waveguide. Considering a system of two  V-type three-level  atoms interacting  via a waveguide, we show that the coherent interference between the spontaneous emission from different excited levels of the two atoms results in a collective quantum beat phenomenon \cite{Han21} (Section~\ref{Sec:model}). We find that  the distance between the atoms modulo the beat wavelength ($d/\lambda_\mr{beat}$) is critical in determining the interference properties of such collective quantum beats: the emitted fields at the two transition frequencies go from interfering constructively to destructively for $ d = n\lambda_\mr{beat}$ to $d = \bkt{n + 1/2}\lambda_\mr{beat}$  (Fig.~\ref{Fig:beatphase}). We obtain the collective dynamics of the total atom+field system by analyzing the system in terms of its characteristic complex eigenfrequenices determined by the poles of the system propagator  (Section~\ref{Sec:CollDyn}). In the limit of coincident atoms ($d\rightarrow0$) our results agree with the recent experimental investigations of vacuum-induced collective quantum beats \cite{Han21}.  A general analysis of the collective atomic and field dynamics as a function of the interatomic separation and the initial atomic states is presented in Section~\ref{Sec:Dist}. We find that  while the atomic separation modulo the transition wavelength in conjunction with the symmetry properties of the initial state governs the overall collective spontaneous emission, the length scale $\lambda_\mr{beat}$ governs the collective nature of the quantum beats. We further investigate the non-Markovian dynamics of collective quantum beats in Section~\ref{Sec:dist_nm}. As the system size become comparable to the coherence length of the emitted photons $(d\sim v/\Gamma)$, there can be significant retardation effects in the field mediating the interaction between the atoms, rendering the system dynamics non-Markovian.  In such a regime, we find that the collective quantum beats can exhibit a non-Markovian enhancement beyond `superradiant' quantum beats arising in the Markovian regime, as illustrated in Fig.~\ref{fig:nonmarkovian_dynamics}.

The results presented in this work open new directions for investigating and controlling radiation from multilevel atomic systems coupled to waveguides. Quantum beats are relevant to precision time-resolved spectroscopy measurements \cite{Haroche1976}; a collective enhancement of quantum beats can improve sensitivities of such measurements. It would be pertinent to extend the present model to a system of $N$ atoms for characterizing the metrological advantage offered by the collective nature of quantum beats.

Furthermore, collections of quantum emitters coupled to waveguides are a paradigmatic system for implementation of quantum networks and long-distance quantum communication protocols. It has been shown that collective multilevel atomic systems offer a higher dimensional entangled state space, enabling efficient quantum memories \cite{Asenjo19} and secure quantum communication \cite{Cerf02, Vaziri02, Tittel2000}. Our analysis is relevant to such schemes,  with a consideration of retardation effects, which can be significant in distributed quantum information processing.

\section{Acknowledgments}
We thank Pablo Solano for insightful discussions.  This research was supported by U.S. Army Research Laboratory's Maryland ARL Quantum Partnership (W911NF-19-2-0181) and Joint Quantum Institute (70NANB16H168). 

\appendix
\begin{widetext}

\section{Atomic dynamics}
\label{App:dyn}

We can solve the coupled atomic equations of motion (Eq.~\eqref{eq:de}) by using Laplace transformation. Defining Laplace-transformed coefficients: $\tilde{c}_{m,j}(s) \equiv \int_0^{\infty} c_{m,j}(t) e^{-st} \dd(t)$, Eq.~\eqref{eq:de} reads
\eqn{
s\tilde{c}_{m,j}(s)-c_{m,j}(0)=-\sum_{n=A,B}\sum_{l=2,3}\frac{\Gamma_{jl}}{2} e^{-\frac{\abs{x_m-x_n}}{v}\bkt{s-i\omega_{j1}}} \tilde{c}_{n,l}(s-i\omega_{jl}).
}
Putting the initial state condition (Eq.~\eqref{eq:psi-0}), we get the coupled equations in Laplace space.
\begin{subequations}
	\begin{align}
	s\tilde{c}_{A2}(s)&=\cos{\theta}-\frac{\Gamma_{22}}{2}\tilde{c}_{A2}(s)-\frac{\Gamma_{23}}{2}\tilde{c}_{A3}(s-i\omega_{23})-\frac{\Gamma_{22}}{2}e^{i\phi_2}e^{-\frac{d}{v}s}\tilde{c}_{B2}(s)-\frac{\Gamma_{23}}{2}e^{i\phi_2}e^{-\frac{d}{v}s} \tilde{c}_{B3}(s-i\omega_{23}),\\
	s\tilde{c}_{B2}(s)&=e^{i\phi}\sin{\theta}-\frac{\Gamma_{22}}{2}e^{i\phi_2}e^{-\frac{d}{v}s}\tilde{c}_{A2}(s)-\frac{\Gamma_{23}}{2}e^{i\phi_2}e^{-\frac{d}{v}s}\tilde{c}_{A3}(s-i\omega_{23})-\frac{\Gamma_{22}}{2}\tilde{c}_{B2}(s)-\frac{\Gamma_{23}}{2} \tilde{c}_{B3}(s-i\omega_{23}),\\	s\tilde{c}_{A3}(s)&=-\frac{\Gamma_{32}}{2}\tilde{c}_{A2}(s+i\omega_{23})-\frac{\Gamma_{33}}{2}\tilde{c}_{A3}(s)-\frac{\Gamma_{32}}{2}e^{i\phi_3}e^{-\frac{d}{v}s}\tilde{c}_{B2}(s+i\omega_{23})-\frac{\Gamma_{33}}{2}e^{i\phi_3}e^{-\frac{d}{v}s} \tilde{c}_{B3}(s),\\	s\tilde{c}_{B3}(s)&=-\frac{\Gamma_{32}}{2}e^{i\phi_3}e^{-\frac{d}{v}s}\tilde{c}_{A2}(s+i\omega_{23})-\frac{\Gamma_{33}}{2}e^{i\phi_3}e^{-\frac{d}{v}s}\tilde{c}_{A3}(s)-\frac{\Gamma_{32}}{2}\tilde{c}_{B2}(s+i\omega_{23})-\frac{\Gamma_{33}}{2} \tilde{c}_{B3}(s).
	\end{align}
\end{subequations}
Here, $\phi_j=\omega_{j1}d/v$ is the general propagation phase for a photon of frequency $\omega_{j1}$. First solving for $\tilde{c}_{A2}(s)$ and $\tilde{c}_{B2}(s)$, we get
\eqn{
\begin{split}
    \tilde{c}_{A2}(s)&=\frac{\cos{\theta}\,A(s)-e^{i\phi}\sin{\theta}\,B(s)}{C(s)},\\
    \tilde{c}_{B2}(s)&=\frac{e^{i\phi}\sin{\theta}\,A(s)-\cos{\theta}\,B(s)}{C(s)},
    \label{eq:c2s}
\end{split}
}
where $A(s)$, $B(s)$, $C(s)$, and $D(s)$ are defined as
\eqn{\begin{split}
    A(s) &= \bkt{s-i\omega_{23}+\frac{\Gamma_{33}}{2}}\sbkt{\bkt{s+\frac{\Gamma_{22}}{2}}\bkt{s-i\omega_{23}+\frac{\Gamma_{33}}{2}}-\frac{\Gamma_{23}\Gamma_{32}}{4}}\\ &\qquad\qquad-e^{-2\frac{d}{v}(s-i\omega_{21})}\sbkt{\bkt{\frac{\Gamma_{33}}{2}}^2\bkt{s+\Gamma_{22}}+\frac{\Gamma_{23}\Gamma_{32}}{4}\bkt{s-i\omega_{23}-\frac{\Gamma_{33}}{2}}},\\
    B(s) &= e^{i\phi_2}e^{-\frac{d}{v}s}\sbkt{\frac{\Gamma_{22}}{2}\bkt{s-i\omega_{23}+\frac{\Gamma_{33}}{2}}^2-\frac{\Gamma_{23}\Gamma_{32}}{4}\bkt{2s-2i\omega_{23}+\frac{\Gamma_{33}}{2}}} -e^{-3\frac{d}{v}(s-i\omega_{21})}\frac{\Gamma_{33}}{2}\sbkt{\frac{\Gamma_{22}\Gamma_{33}}{4}-\frac{\Gamma_{23}\Gamma_{32}}{4}},\\
    C(s) &= \sbkt{\bkt{s-i\omega_{23}+\frac{\Gamma_{33}}{2}+\frac{\Gamma_{33}}{2} e^{i\phi_2}e^{-\frac{d}{v}s}} \bkt{s+\frac{\Gamma_{22}}{2}+\frac{\Gamma_{22}}{2} e^{i\phi_2}e^{-\frac{d}{v}s}}-\frac{\Gamma_{23}\Gamma_{32}}{4}\bkt{1+e^{i\phi_2}e^{-\frac{d}{v}s}}^2}\\
    &\quad \times \sbkt{\bkt{s-i\omega_{23}+\frac{\Gamma_{33}}{2}-\frac{\Gamma_{33}}{2} e^{i\phi_2}e^{-\frac{d}{v}s}} \bkt{s+\frac{\Gamma_{22}}{2}-\frac{\Gamma_{22}}{2} e^{i\phi_2}e^{-\frac{d}{v}s}}-\frac{\Gamma_{23}\Gamma_{32}}{4}\bkt{1-e^{i\phi_2}e^{-\frac{d}{v}s}}^2}.
    \label{eq:ABCD}
\end{split}}

Then $\tilde{c}_{A3}(s)$ and $\tilde{c}_{B3}(s)$ are obtained in terms of $\tilde{c}_{A2}(s)$ and $\tilde{c}_{B2}(s)$:
\eqn{
\begin{split}
    \tilde{c}_{A3}(s)&=-\frac{\Gamma_{32}}{2}
    \frac{
    \bkt{s+\frac{\Gamma_{33}}{2}-\frac{\Gamma_{33}}{2}e^{-2\frac{d}{v}(s-i\omega_{31})}} \tilde{c}_{A2}(s+i\omega_{23}) + s e^{i\phi_3}e^{-\frac{d}{v}s} \tilde{c}_{B2}(s+i\omega_{23}) }{\bkt{s+\frac{\Gamma_{33}}{2}}^2-\bkt{\frac{\Gamma_{33}}{2}}^2 e^{-2\frac{d}{v}(s-i\omega_{31})}},\\
    \tilde{c}_{B3}(s)&=-\frac{\Gamma_{32}}{2}
    \frac{
     s e^{i\phi_3}e^{-\frac{d}{v}s}\tilde{c}_{A2}(s+i\omega_{23}) + \bkt{s+\frac{\Gamma_{33}}{2}-\frac{\Gamma_{33}}{2}e^{-2\frac{d}{v}(s-i\omega_{31})}}\tilde{c}_{B2}(s+i\omega_{23}) }{\bkt{s+\frac{\Gamma_{33}}{2}}^2-\bkt{\frac{\Gamma_{33}}{2}}^2 e^{-2\frac{d}{v}(s-i\omega_{31})}},
    \label{eq:c3s}
\end{split}
}

We first numerically find the poles $s_n$ of the denominators in Eqs.~\eqref{eq:c2s} and \eqref{eq:c3s}, with  each pole representing a complex eigenfrequency of the system.  The excitation amplitude $\tilde{c}(s)$ in Laplace space is expressed in terms of its modes:

\eqn{
    \tilde{c}(s)&=\sum_n\frac{w_n}{s-s_n},
    \label{eq:solution-laplace}
}
where $w_n =\lim_{s \to s_n} (s-s_n) \tilde{c}(s)$.

In this paper we consider two specific initial states: a symmetric and anti-symmetric superposition of a single excitation in level 2. For the symmetric initial state, i.e., $\theta=\pi/4$ and $\phi=0$,
\eqn{
    \label{eq:symmetric-solution-2}
    \tilde{c}_{A2}(s) = \tilde{c}_{B2}(s) &= \frac{1}{\sqrt{2}} \frac{s-i\omega_{23}+\frac{\Gamma_{33}}{2}+\frac{\Gamma_{33}}{2}e^{i\phi_2}e^{-\frac{d}{v}s}}{(s-i\omega_{23}+\frac{\Gamma_{33}}{2}+\frac{\Gamma_{33}}{2}e^{i\phi_2}e^{-\frac{d}{v}s})(s+\frac{\Gamma_{22}}{2}+\frac{\Gamma_{22}}{2}e^{i\phi_2}e^{-\frac{d}{v}s})-\frac{\Gamma_{23}\Gamma_{32}}{4}(1+e^{i\phi_2}e^{-\frac{d}{v}s})^2},\\
    \label{eq:symmetric-solution-3}
    \tilde{c}_{A3}(s) = \tilde{c}_{B3}(s) &= -\frac{\Gamma_{32}}{2\sqrt{2}}\frac{1+e^{i\phi_3}e^{-\frac{d}{v}s}}{(s+\frac{\Gamma_{33}}{2}+\frac{\Gamma_{33}}{2}e^{i\phi_3}e^{-\frac{d}{v}s})(s+i\omega_{23}+\frac{\Gamma_{22}}{2}+\frac{\Gamma_{22}}{2}e^{i\phi_3}e^{-\frac{d}{v}s})-\frac{\Gamma_{23}\Gamma_{32}}{4}(1+e^{i\phi_3}e^{-\frac{d}{v}s})^2}.
}
Note that the denominators in Eqs.~\eqref{eq:symmetric-solution-2} and \eqref{eq:symmetric-solution-3}  are the same up to a constant shift of the Laplace variable $s\rightarrow s + i\omega_{23}$.

\begin{figure}[]
    %\raggedleft
    \centering
    \includegraphics[width = 5.5 in]{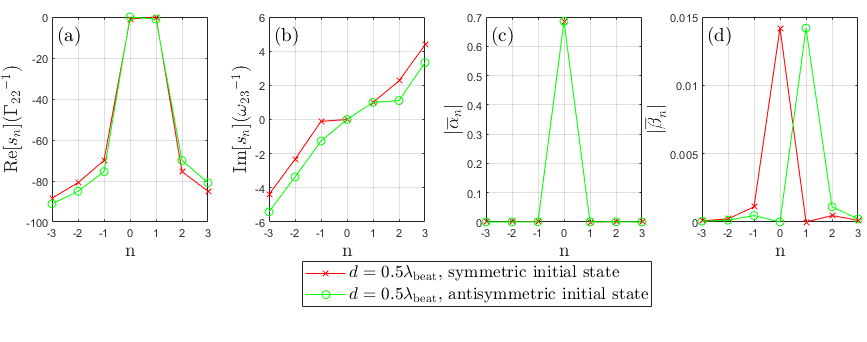}
    \includegraphics[width = 5.5 in]{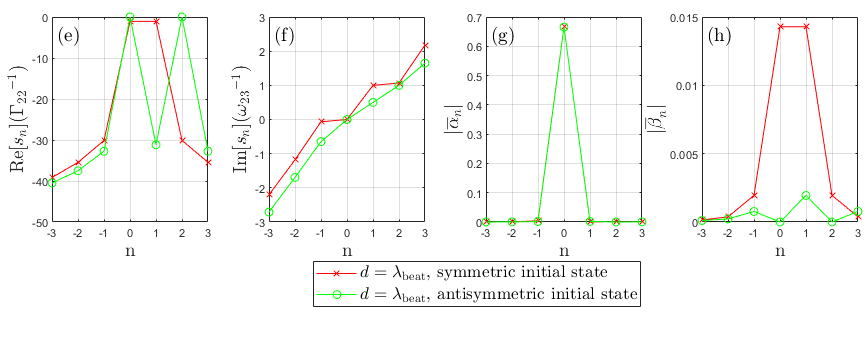}
    \caption{
    Markovian regime: the poles and the corresponding coefficients for symmetric(red x) and the anti-symmetric(green circle) initial states, for the separation of (a-d) $d=0.5\,\lambda_\mr{beat}$ and (e-h) $d=\lambda_\mr{beat}$. (a,e) The real part of the poles corresponds to the decay rate of each modes and (b,f) the imaginary part to the frequency of the modes. (c,g) The coefficient $\overline{\alpha}_n$ shows the contribution of each pole to $\ket{2}$ dynamics and (d,h) the coefficient $\overline{\beta}_n$ shows the contribution of each pole to $\ket{3}$ dynamics. 
    }
    \label{Fig:poles_m}
\end{figure}

Similarly, for  an anti-symmetric initial state, i.e., $\theta=\pi/4$ and $\phi=\pi$,
\eqn{
   \label{eq:anti-symmetric-solution-2}
   \tilde{c}_{A2}(s) = -\tilde{c}_{B2}(s) &= \frac{1}{\sqrt{2}} \frac{s-i\omega_{23}+\frac{\Gamma_{33}}{2}-\frac{\Gamma_{33}}{2}e^{i\phi_2}e^{-\frac{d}{v}s}}{(s-i\omega_{23}+\frac{\Gamma_{33}}{2}-\frac{\Gamma_{33}}{2}e^{i\phi_2}e^{-\frac{d}{v}s})(s+\frac{\Gamma_{22}}{2}-\frac{\Gamma_{22}}{2}e^{i\phi_2}e^{-\frac{d}{v}s})-\frac{\Gamma_{23}\Gamma_{32}}{4}(1-e^{i\phi_2}e^{-\frac{d}{v}s})^2},\\
   \label{eq:anti-symmetric-solution-3}
   \tilde{c}_{A3}(s) = -\tilde{c}_{B3}(s) &= -\frac{\Gamma_{32}}{2\sqrt{2}}\frac{1-e^{i\phi_3}e^{-\frac{d}{v}s}}{(s+\frac{\Gamma_{33}}{2}-\frac{\Gamma_{33}}{2}e^{i\phi_3}e^{-\frac{d}{v}s})(s+i\omega_{23}+\frac{\Gamma_{22}}{2}-\frac{\Gamma_{22}}{2}e^{i\phi_3}e^{-\frac{d}{v}s})-\frac{\Gamma_{23}\Gamma_{32}}{4}(1-e^{i\phi_3}e^{-\frac{d}{v}s})^2}.
}
Similar to the symmetric case,  the denominators on the RHS of    Eq.~\eqref{eq:anti-symmetric-solution-2} and    \label{eq:anti-symmetric-solution-1} are the same up to a Laplace variable shift of  $s\rightarrow s + i\omega_{23}$ as well. 

We remark that the dynamics of a general initial state in the single excitation manifold $\ket{\Psi(0)} =  K_{2A}\ket{2}_A\ket{1}_B +K_{2B}\ket{1}_A\ket{2}_B +K_{3A}\ket{3}_A\ket{1}_B +K_{3B}\ket{1}_A\ket{3}_B,$ with $\abs{K_{2A}}^2+\abs{K_{2B}}^2+\abs{K_{3A}}^2+\abs{K_{3B}}^2=1$ can be obtained directly from our result. An initial excitation in $\ket{3}_{A} (\ket{3}_B)$ would follow the same dynamics as for the case of an initial excitation in $\ket{2}_A (\ket{2}_B)$,  given by Eqs.~\eqref{eq:c2t}  and \eqref{eq:c3t}, with the following substitutions:
\eqn{
  \omega_{23} &\leftrightarrow -\omega_{23}\nonumber\\
  \Gamma_{22} &\leftrightarrow \Gamma_{33}\nonumber\\
  \Gamma_{23} &\leftrightarrow \Gamma_{32}\nonumber.
}

\section{Field Intensity dynamics}
\label{App:Int}
We can derive the excitation amplitudes of the field modes from Eq.~\eqref{eq:EOMa} and \eqref{eq:EOMb} as follows:
\eqn{\label{Eq:ca}
c_R \bkt{\omega_k , t} =
	 &i\int_{0}^{t}\dd{\tau}\sum_{m=A,B}\sum_{j=2,3}g_j(\omega_k)e^{-i(\omega_{j1}-\omega_{k})\tau}   c_{m,j}(\tau) e^{-ik\cdot x_m} \\
\label{Eq:cb}
c_L \bkt{\omega_k , t} =  &i\int_{0}^{t}\dd{\tau}\sum_{m=A,B}\sum_{j=2,3}g_j(\omega_k)e^{-i(\omega_{j1}-\omega_{k})\tau}   c_{m,j}(\tau) e^{ik\cdot x_m} }

We consider the dynamics of the intensity radiated by the atoms as follows:

\eqn{
&I\bkt{x,t}\non\\
&=\frac{\epsilon_0 c}{2}\bra{\Psi\bkt{t}} \sbkt{\int _0 ^\infty\dd k_1 \mc{E}_{k_1} ^\ast\cbkt{ \hat{a}_{R,k_1}^\dagger e^{-ik_1x}+ \hat{a}_{L,k_1}^\dagger e^{ik_1 x} }e^{i \omega_1 t}}\sbkt{\int _0 ^\infty\dd k_2 \mc{E}_{k_2} \cbkt{ \hat{a}_{R,k_2}e^{ik_2x}+ \hat{a}_{L,k_2}e^{-ik_2x} }e^{-i \omega_2 t}} \ket{\Psi\bkt{t}}\\
&= 
\frac{\epsilon_0 c\abs{\mc{E}_0}^2}{2} \abs{\int_0 ^\infty \dd k \sbkt{c_R\bkt{\omega_k , t}e^{i k x} + c_L\bkt{\omega_k , t}e^{-i k x} } e^{-i \omega_k t}}^2.
}
where we have assumed that $\mc{E}_k\approx\mc{E}_{0}$ to be constant near the atomic resonance frequency.

One can simplify the above in terms of the atomic coefficients as follows:
\eqn{
I\bkt{x,t} = \frac{\epsilon_0 c \abs{\mc{E}_0}^2 \beta}{4\pi}&\abs{\int \dd\omega e^{-i\omega t} \sbkt{ \int_0 ^t \dd\tau\sum_{j = 2,3} g_j \cbkt{ c_{Aj} \bkt{\tau } e^{i \omega \bkt{- x + x_A}/v}+ c_{Bj} \bkt{\tau } e^{i \omega \bkt{- x + x_B}/v} \right.\right.\right.\non\\
&\left.\left.\left.+ c_{Aj} \bkt{\tau } e^{-i \omega \bkt{- x + x_A}/v}+ c_{Bj} \bkt{\tau } e^{-i \omega \bkt{- x + x_B}/v}} e^{i \bkt{\omega - \omega_{j1}} \tau}} }^2,
}
where we have used Eq.~\eqref{Eq:ca} and \eqref{Eq:cb} to determine the field excitation amplitudes in terms of those of the atoms. Performing first the integral over frequency and subsequently the integral over time leads to Eq.~\eqref{Eq:int}.

\section{Poles for the Markovian and the Non-Markovian examples}
\label{App:poles}

\begin{figure}[]
    %\raggedleft
    \centering
    \includegraphics[width = 5.5 in]{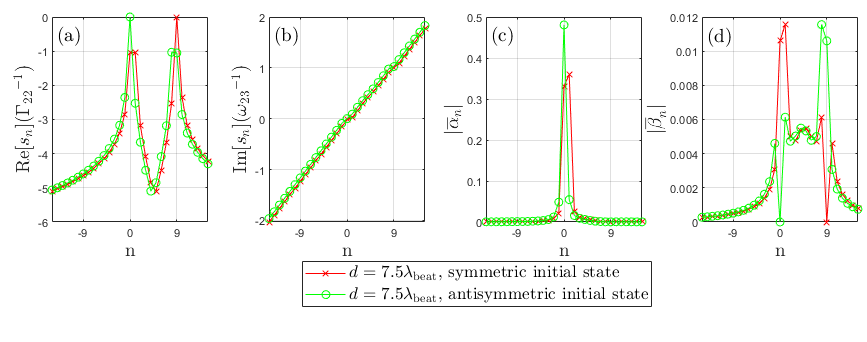}
    \includegraphics[width = 5.5 in]{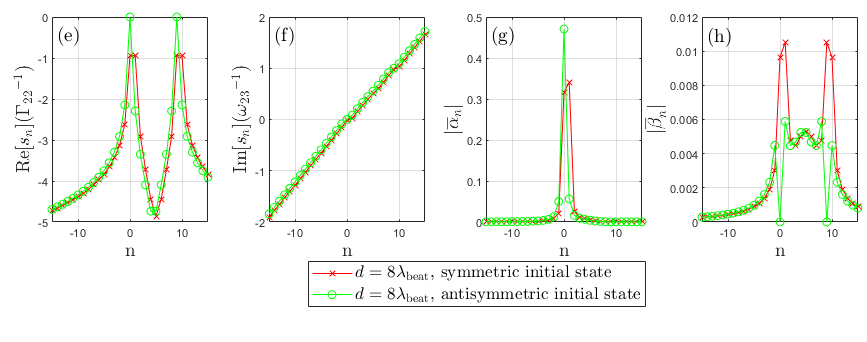}
    \caption{
    Non-Markovian regime: The example of the poles and the corresponding coefficients for symmetric(red x) and the anti-symmetric(green circle) initial states, for (a-d) $d=7.5\,\lambda_\mr{beat}$ and (e-h) $d=8\lambda_\mr{beat}$. (a,e) The real part of the poles corresponds to the decay rate of each mode and (b,f) the imaginary part corresponds to their frequency. (c,g) The coefficient $\overline{\alpha}_n$ represents the contribution of each pole to the $\ket{2}$ dynamics and (d,h)  $\overline{\beta}_n$ to that of level  $\ket{3}$.
    }
    \label{Fig:poles_nm}
\end{figure}

The poles $s_n$ and the corresponding coefficients $\alpha_n^{(\pm)}$ and $\beta_n^{(\pm)}$  in Eq.~\eqref{eq:c2t} and \eqref{eq:c3t}  determine the atomic and the field dynamics. However, the coefficients  $\alpha_n^{(\pm)}$ and $\beta_n^{(\pm)}$ also account for the higher eigenfrequencies that cause an abrupt change at $t=d/v$, arising due to the time-delayed feedback. Since we are interested in the dynamics after $t=d/v$, we redefine the following coefficients:
\eqn{
\overline{\alpha}_n \equiv \alpha_n e^{s_n d/v},\\
\overline{\beta}_n \equiv \beta_n e^{s_n d/v}.\\
}
The  dynamics after $t=d/v$ is thus described as
\eqn{
\label{dyn:c2t-new}
c_{A2}\bkt{t} &= \sum_{n = -\infty}^{\infty} \overline{\alpha}_n e^{s_n (t-d/v)},\\
\label{dyn:c3t-new}
c_{A3}\bkt{t} &= \sum_{n = -\infty}^{\infty} \overline{\beta}_n e^{\bkt{s_n-i\omega_{23}} (t-d/v)}.
}
$\overline{\alpha}_n$ and $\overline{\beta}_n$ do not consider the high frequency components constituting the abrupt change at $t=d/v$, and only account for the dynamics in the regime $t>d/v$.

Fig.~\ref{Fig:poles_m} and \ref{Fig:poles_nm} show the poles $s_n$ and the corresponding coefficients $\overline{\alpha}_n$ and $\overline{\beta}_n$ for Markovian and non-Markovian regimes, respectively. The real part of the poles represents the collective decay rate and the imaginary part represents collective shift of energy in the eigenmodes. We see that in Markovian regime (Fig.~\ref{Fig:poles_m}) the high frequency and fast decaying modes have negligible contribution. In contrast, the non-Markovian regime (Fig.~\ref{Fig:poles_nm}) shows that the dominant poles occur in the range of frequencies between 0 and $\omega_{23}$, and decay rates that are enhanced beyond Markovian limit.

\end{widetext}

\bibliography{WGQED}

\begin{thebibliography}{43}%
\makeatletter
\providecommand \@ifxundefined [1]{%
 \@ifx{#1\undefined}
}%
\providecommand \@ifnum [1]{%
 \ifnum #1\expandafter \@firstoftwo
 \else \expandafter \@secondoftwo
 \fi
}%
\providecommand \@ifx [1]{%
 \ifx #1\expandafter \@firstoftwo
 \else \expandafter \@secondoftwo
 \fi
}%
\providecommand \natexlab [1]{#1}%
\providecommand \enquote  [1]{``#1''}%
\providecommand \bibnamefont  [1]{#1}%
\providecommand \bibfnamefont [1]{#1}%
\providecommand \citenamefont [1]{#1}%
\providecommand \href@noop [0]{\@secondoftwo}%
\providecommand \href [0]{\begingroup \@sanitize@url \@href}%
\providecommand \@href[1]{\@@startlink{#1}\@@href}%
\providecommand \@@href[1]{\endgroup#1\@@endlink}%
\providecommand \@sanitize@url [0]{\catcode `\\12\catcode `\$12\catcode
  `\&12\catcode `\#12\catcode `\^12\catcode `\_12\catcode `\%12\relax}%
\providecommand \@@startlink[1]{}%
\providecommand \@@endlink[0]{}%
\providecommand \url  [0]{\begingroup\@sanitize@url \@url }%
\providecommand \@url [1]{\endgroup\@href {#1}{\urlprefix }}%
\providecommand \urlprefix  [0]{URL }%
\providecommand \Eprint [0]{\href }%
\providecommand \doibase [0]{http://dx.doi.org/}%
\providecommand \selectlanguage [0]{\@gobble}%
\providecommand \bibinfo  [0]{\@secondoftwo}%
\providecommand \bibfield  [0]{\@secondoftwo}%
\providecommand \translation [1]{[#1]}%
\providecommand \BibitemOpen [0]{}%
\providecommand \bibitemStop [0]{}%
\providecommand \bibitemNoStop [0]{.\EOS\space}%
\providecommand \EOS [0]{\spacefactor3000\relax}%
\providecommand \BibitemShut  [1]{\csname bibitem#1\endcsname}%
\let\auto@bib@innerbib\@empty
%</preamble>
\bibitem [{\citenamefont {Jaynes~in}(1980)}]{Jaynes_1980}%
  \BibitemOpen
  \bibfield  {author} {\bibinfo {author} {\bibfnamefont {E.~T.}\ \bibnamefont
  {Jaynes~in}},\ }\href {\doibase 10.1007/978-1-4757-0671-0_3} {\emph {\bibinfo
  {title} {Foundations of Radiation Theory and Quantum Electrodynamics}}},\
  edited by\ \bibinfo {editor} {\bibfnamefont {A.~O.}\ \bibnamefont {Barut}}\
  (\bibinfo  {publisher} {Springer US},\ \bibinfo {address} {Boston, MA},\
  \bibinfo {year} {1980})\ pp.\ \bibinfo {pages} {37--43}\BibitemShut {NoStop}%
\bibitem [{\citenamefont {Dicke}(1954)}]{Dicke_1954}%
  \BibitemOpen
  \bibfield  {author} {\bibinfo {author} {\bibfnamefont {R.~H.}\ \bibnamefont
  {Dicke}},\ }\href {\doibase 10.1103/PhysRev.93.99} {\bibfield  {journal}
  {\bibinfo  {journal} {Phys. Rev.}\ }\textbf {\bibinfo {volume} {93}},\
  \bibinfo {pages} {99} (\bibinfo {year} {1954})}\BibitemShut {NoStop}%
\bibitem [{\citenamefont {Agarwal}(1977)}]{Agarwal_1977}%
  \BibitemOpen
  \bibfield  {author} {\bibinfo {author} {\bibfnamefont {G.~S.}\ \bibnamefont
  {Agarwal}},\ }\href {\doibase 10.1103/PhysRevA.15.2380} {\bibfield  {journal}
  {\bibinfo  {journal} {Phys. Rev. A}\ }\textbf {\bibinfo {volume} {15}},\
  \bibinfo {pages} {2380} (\bibinfo {year} {1977})}\BibitemShut {NoStop}%
\bibitem [{\citenamefont {Han}\ \emph {et~al.}(2021)\citenamefont {Han},
  \citenamefont {Lee}, \citenamefont {Sinha}, \citenamefont {Fatemi},\ and\
  \citenamefont {Rolston}}]{Han21}%
  \BibitemOpen
  \bibfield  {author} {\bibinfo {author} {\bibfnamefont {H.~S.}\ \bibnamefont
  {Han}}, \bibinfo {author} {\bibfnamefont {A.}~\bibnamefont {Lee}}, \bibinfo
  {author} {\bibfnamefont {K.}~\bibnamefont {Sinha}}, \bibinfo {author}
  {\bibfnamefont {F.~K.}\ \bibnamefont {Fatemi}}, \ and\ \bibinfo {author}
  {\bibfnamefont {S.~L.}\ \bibnamefont {Rolston}},\ }\href {\doibase
  10.1103/PhysRevLett.127.073604} {\bibfield  {journal} {\bibinfo  {journal}
  {Phys. Rev. Lett.}\ }\textbf {\bibinfo {volume} {127}},\ \bibinfo {pages}
  {073604} (\bibinfo {year} {2021})}\BibitemShut {NoStop}%
\bibitem [{\citenamefont {Haroche}(1976)}]{Haroche1976}%
  \BibitemOpen
  \bibfield  {author} {\bibinfo {author} {\bibfnamefont {S.}~\bibnamefont
  {Haroche}},\ }\enquote {\bibinfo {title} {Quantum beats and time-resolved
  fluorescence spectroscopy},}\ in\ \href {\doibase 10.1007/3540077197_23}
  {\emph {\bibinfo {booktitle} {High-Resolution Laser Spectroscopy}}},\
  \bibinfo {editor} {edited by\ \bibinfo {editor} {\bibfnamefont
  {K.}~\bibnamefont {Shimoda}}}\ (\bibinfo  {publisher} {Springer Berlin
  Heidelberg},\ \bibinfo {address} {Berlin, Heidelberg},\ \bibinfo {year}
  {1976})\ pp.\ \bibinfo {pages} {253--313}\BibitemShut {NoStop}%
\bibitem [{\citenamefont {Gross}\ \emph {et~al.}(1976)\citenamefont {Gross},
  \citenamefont {Fabre}, \citenamefont {Pillet},\ and\ \citenamefont
  {Haroche}}]{Gross76}%
  \BibitemOpen
  \bibfield  {author} {\bibinfo {author} {\bibfnamefont {M.}~\bibnamefont
  {Gross}}, \bibinfo {author} {\bibfnamefont {C.}~\bibnamefont {Fabre}},
  \bibinfo {author} {\bibfnamefont {P.}~\bibnamefont {Pillet}}, \ and\ \bibinfo
  {author} {\bibfnamefont {S.}~\bibnamefont {Haroche}},\ }\href {\doibase
  10.1103/PhysRevLett.36.1035} {\bibfield  {journal} {\bibinfo  {journal}
  {Phys. Rev. Lett.}\ }\textbf {\bibinfo {volume} {36}},\ \bibinfo {pages}
  {1035} (\bibinfo {year} {1976})}\BibitemShut {NoStop}%
\bibitem [{\citenamefont {Gross}\ and\ \citenamefont
  {Haroche}(1982)}]{Gross82}%
  \BibitemOpen
  \bibfield  {author} {\bibinfo {author} {\bibfnamefont {M.}~\bibnamefont
  {Gross}}\ and\ \bibinfo {author} {\bibfnamefont {S.}~\bibnamefont
  {Haroche}},\ }\href {https://doi.org/10.1016/0370-1573(82)90102-8} {\bibfield
   {journal} {\bibinfo  {journal} {Phys. Rep.}\ }\textbf {\bibinfo {volume}
  {93}},\ \bibinfo {pages} {301} (\bibinfo {year} {1982})}\BibitemShut
  {NoStop}%
\bibitem [{\citenamefont {Skribanowitz}\ \emph {et~al.}(1973)\citenamefont
  {Skribanowitz}, \citenamefont {Herman}, \citenamefont {MacGillivray},\ and\
  \citenamefont {Feld}}]{Skribanowitz73}%
  \BibitemOpen
  \bibfield  {author} {\bibinfo {author} {\bibfnamefont {N.}~\bibnamefont
  {Skribanowitz}}, \bibinfo {author} {\bibfnamefont {I.~P.}\ \bibnamefont
  {Herman}}, \bibinfo {author} {\bibfnamefont {J.~C.}\ \bibnamefont
  {MacGillivray}}, \ and\ \bibinfo {author} {\bibfnamefont {M.~S.}\
  \bibnamefont {Feld}},\ }\href {\doibase 10.1103/PhysRevLett.30.309}
  {\bibfield  {journal} {\bibinfo  {journal} {Phys. Rev. Lett.}\ }\textbf
  {\bibinfo {volume} {30}},\ \bibinfo {pages} {309} (\bibinfo {year}
  {1973})}\BibitemShut {NoStop}%
\bibitem [{\citenamefont {Pavolini}\ \emph {et~al.}(1985)\citenamefont
  {Pavolini}, \citenamefont {Crubellier}, \citenamefont {Pillet}, \citenamefont
  {Cabaret},\ and\ \citenamefont {Liberman}}]{Pavolini85}%
  \BibitemOpen
  \bibfield  {author} {\bibinfo {author} {\bibfnamefont {D.}~\bibnamefont
  {Pavolini}}, \bibinfo {author} {\bibfnamefont {A.}~\bibnamefont
  {Crubellier}}, \bibinfo {author} {\bibfnamefont {P.}~\bibnamefont {Pillet}},
  \bibinfo {author} {\bibfnamefont {L.}~\bibnamefont {Cabaret}}, \ and\
  \bibinfo {author} {\bibfnamefont {S.}~\bibnamefont {Liberman}},\ }\href
  {\doibase 10.1103/PhysRevLett.54.1917} {\bibfield  {journal} {\bibinfo
  {journal} {Phys. Rev. Lett}\ }\textbf {\bibinfo {volume} {54}},\ \bibinfo
  {pages} {1917} (\bibinfo {year} {1985})}\BibitemShut {NoStop}%
\bibitem [{\citenamefont {DeVoe}\ and\ \citenamefont {Brewer}(1996)}]{Devoe96}%
  \BibitemOpen
  \bibfield  {author} {\bibinfo {author} {\bibfnamefont {R.~G.}\ \bibnamefont
  {DeVoe}}\ and\ \bibinfo {author} {\bibfnamefont {R.~G.}\ \bibnamefont
  {Brewer}},\ }\href {\doibase 10.1103/PhysRevLett.76.2049} {\bibfield
  {journal} {\bibinfo  {journal} {Phys. Rev. Lett.}\ }\textbf {\bibinfo
  {volume} {76}},\ \bibinfo {pages} {2049} (\bibinfo {year}
  {1996})}\BibitemShut {NoStop}%
\bibitem [{\citenamefont {Sheremet}\ \emph {et~al.}(2021)\citenamefont
  {Sheremet}, \citenamefont {Petrov}, \citenamefont {Iorsh}, \citenamefont
  {Poshakinskiy},\ and\ \citenamefont {Poddubny}}]{Sheremet2021WaveguideQE}%
  \BibitemOpen
  \bibfield  {author} {\bibinfo {author} {\bibfnamefont {A.~S.}\ \bibnamefont
  {Sheremet}}, \bibinfo {author} {\bibfnamefont {M.~I.}\ \bibnamefont
  {Petrov}}, \bibinfo {author} {\bibfnamefont {I.~V.}\ \bibnamefont {Iorsh}},
  \bibinfo {author} {\bibfnamefont {A.~V.}\ \bibnamefont {Poshakinskiy}}, \
  and\ \bibinfo {author} {\bibfnamefont {A.~N.}\ \bibnamefont {Poddubny}}\
  }(\bibinfo {year} {2021})\ \Eprint {http://arxiv.org/abs/2103.06824}
  {arXiv:2103.06824 [quant-ph]} \BibitemShut {NoStop}%
\bibitem [{\citenamefont {Pivovarov}\ \emph {et~al.}(2021)\citenamefont
  {Pivovarov}, \citenamefont {Gerasimov}, \citenamefont {Berroir},
  \citenamefont {Ray}, \citenamefont {Laurat}, \citenamefont {Urvoy},\ and\
  \citenamefont {Kupriyanov}}]{Pivovarov21}%
  \BibitemOpen
  \bibfield  {author} {\bibinfo {author} {\bibfnamefont {V.~A.}\ \bibnamefont
  {Pivovarov}}, \bibinfo {author} {\bibfnamefont {L.~V.}\ \bibnamefont
  {Gerasimov}}, \bibinfo {author} {\bibfnamefont {J.}~\bibnamefont {Berroir}},
  \bibinfo {author} {\bibfnamefont {T.}~\bibnamefont {Ray}}, \bibinfo {author}
  {\bibfnamefont {J.}~\bibnamefont {Laurat}}, \bibinfo {author} {\bibfnamefont
  {A.}~\bibnamefont {Urvoy}}, \ and\ \bibinfo {author} {\bibfnamefont {D.~V.}\
  \bibnamefont {Kupriyanov}},\ }\href {\doibase 10.1103/PhysRevA.103.043716}
  {\bibfield  {journal} {\bibinfo  {journal} {Phys. Rev. A}\ }\textbf {\bibinfo
  {volume} {103}},\ \bibinfo {pages} {043716} (\bibinfo {year}
  {2021})}\BibitemShut {NoStop}%
\bibitem [{\citenamefont {Asenjo-Garcia}\ \emph {et~al.}(2017)\citenamefont
  {Asenjo-Garcia}, \citenamefont {Moreno-Cardoner}, \citenamefont {Albrecht},
  \citenamefont {Kimble},\ and\ \citenamefont {Chang}}]{Asenjo17}%
  \BibitemOpen
  \bibfield  {author} {\bibinfo {author} {\bibfnamefont {A.}~\bibnamefont
  {Asenjo-Garcia}}, \bibinfo {author} {\bibfnamefont {M.}~\bibnamefont
  {Moreno-Cardoner}}, \bibinfo {author} {\bibfnamefont {A.}~\bibnamefont
  {Albrecht}}, \bibinfo {author} {\bibfnamefont {H.~J.}\ \bibnamefont
  {Kimble}}, \ and\ \bibinfo {author} {\bibfnamefont {D.~E.}\ \bibnamefont
  {Chang}},\ }\href {\doibase 10.1103/PhysRevX.7.031024} {\bibfield  {journal}
  {\bibinfo  {journal} {Phys. Rev. X}\ }\textbf {\bibinfo {volume} {7}},\
  \bibinfo {pages} {031024} (\bibinfo {year} {2017})}\BibitemShut {NoStop}%
\bibitem [{\citenamefont {van Loo}\ \emph {et~al.}(2013)\citenamefont {van
  Loo}, \citenamefont {Fedorov}, \citenamefont {Lalumi\'{e}re}, \citenamefont
  {Sanders}, \citenamefont {Blais},\ and\ \citenamefont
  {Wallraff}}]{vanLoo2013}%
  \BibitemOpen
  \bibfield  {author} {\bibinfo {author} {\bibfnamefont {A.~F.}\ \bibnamefont
  {van Loo}}, \bibinfo {author} {\bibfnamefont {A.}~\bibnamefont {Fedorov}},
  \bibinfo {author} {\bibfnamefont {K.}~\bibnamefont {Lalumi\'{e}re}}, \bibinfo
  {author} {\bibfnamefont {B.~C.}\ \bibnamefont {Sanders}}, \bibinfo {author}
  {\bibfnamefont {A.}~\bibnamefont {Blais}}, \ and\ \bibinfo {author}
  {\bibfnamefont {A.}~\bibnamefont {Wallraff}},\ }\href {\doibase
  10.1126/science.1244324} {\bibfield  {journal} {\bibinfo  {journal}
  {Science}\ }\textbf {\bibinfo {volume} {342}},\ \bibinfo {pages} {1494}
  (\bibinfo {year} {2013})}\BibitemShut {NoStop}%
\bibitem [{\citenamefont {Li}\ and\ \citenamefont {Argyropoulos}(2016)}]{Li16}%
  \BibitemOpen
  \bibfield  {author} {\bibinfo {author} {\bibfnamefont {Y.}~\bibnamefont
  {Li}}\ and\ \bibinfo {author} {\bibfnamefont {C.}~\bibnamefont
  {Argyropoulos}},\ }\href@noop {} {\bibfield  {journal} {\bibinfo  {journal}
  {Opt. Express}\ }\textbf {\bibinfo {volume} {24}},\ \bibinfo {pages} {26696}
  (\bibinfo {year} {2016})}\BibitemShut {NoStop}%
\bibitem [{\citenamefont {Solano}\ \emph
  {et~al.}(2017{\natexlab{a}})\citenamefont {Solano}, \citenamefont
  {Barberis-Blostein}, \citenamefont {Fatemi}, \citenamefont {Orozco},\ and\
  \citenamefont {Rolston}}]{Solano2017}%
  \BibitemOpen
  \bibfield  {author} {\bibinfo {author} {\bibfnamefont {P.}~\bibnamefont
  {Solano}}, \bibinfo {author} {\bibfnamefont {P.}~\bibnamefont
  {Barberis-Blostein}}, \bibinfo {author} {\bibfnamefont {F.~K.}\ \bibnamefont
  {Fatemi}}, \bibinfo {author} {\bibfnamefont {L.~A.}\ \bibnamefont {Orozco}},
  \ and\ \bibinfo {author} {\bibfnamefont {S.~L.}\ \bibnamefont {Rolston}},\
  }\href {\doibase 10.1038/s41467-017-01994-3} {\bibfield  {journal} {\bibinfo
  {journal} {Nat. Commun.}\ }\textbf {\bibinfo {volume} {8}},\ \bibinfo {pages}
  {1857} (\bibinfo {year} {2017}{\natexlab{a}})}\BibitemShut {NoStop}%
\bibitem [{\citenamefont {Kim}\ \emph {et~al.}(2018)\citenamefont {Kim},
  \citenamefont {Aghaeimeibodi}, \citenamefont {Richardson}, \citenamefont
  {Leavitt},\ and\ \citenamefont {Waks}}]{Kim2018}%
  \BibitemOpen
  \bibfield  {author} {\bibinfo {author} {\bibfnamefont {J.-H.}\ \bibnamefont
  {Kim}}, \bibinfo {author} {\bibfnamefont {S.}~\bibnamefont {Aghaeimeibodi}},
  \bibinfo {author} {\bibfnamefont {C.~J.~K.}\ \bibnamefont {Richardson}},
  \bibinfo {author} {\bibfnamefont {R.~P.}\ \bibnamefont {Leavitt}}, \ and\
  \bibinfo {author} {\bibfnamefont {E.}~\bibnamefont {Waks}},\ }\href
  {https://doi.org/10.1021/acs.nanolett.8b01133} {\bibfield  {journal}
  {\bibinfo  {journal} {Nano Letters}\ }\textbf {\bibinfo {volume} {18}},\
  \bibinfo {pages} {4734} (\bibinfo {year} {2018})}\BibitemShut {NoStop}%
\bibitem [{\citenamefont {Newman}\ \emph {et~al.}(2018)\citenamefont {Newman},
  \citenamefont {Cortes}, \citenamefont {Afshar}, \citenamefont {Cadien},
  \citenamefont {Meldrum}, \citenamefont {Fedosejevs},\ and\ \citenamefont
  {Jacob}}]{Newman2018}%
  \BibitemOpen
  \bibfield  {author} {\bibinfo {author} {\bibfnamefont {W.~D.}\ \bibnamefont
  {Newman}}, \bibinfo {author} {\bibfnamefont {C.~L.}\ \bibnamefont {Cortes}},
  \bibinfo {author} {\bibfnamefont {A.}~\bibnamefont {Afshar}}, \bibinfo
  {author} {\bibfnamefont {K.}~\bibnamefont {Cadien}}, \bibinfo {author}
  {\bibfnamefont {A.}~\bibnamefont {Meldrum}}, \bibinfo {author} {\bibfnamefont
  {R.}~\bibnamefont {Fedosejevs}}, \ and\ \bibinfo {author} {\bibfnamefont
  {Z.}~\bibnamefont {Jacob}},\ }\href {\doibase 10.1126/sciadv.aar5278}
  {\bibfield  {journal} {\bibinfo  {journal} {Science Advances}\ }\textbf
  {\bibinfo {volume} {4}},\ \bibinfo {pages} {5278} (\bibinfo {year}
  {2018})}\BibitemShut {NoStop}%
\bibitem [{\citenamefont {Boddeti}\ \emph {et~al.}(2022)\citenamefont
  {Boddeti}, \citenamefont {Guan}, \citenamefont {Sentz}, \citenamefont
  {Juarez}, \citenamefont {Newman}, \citenamefont {Cortes}, \citenamefont
  {Odom},\ and\ \citenamefont {Jacob}}]{Boddeti2022}%
  \BibitemOpen
  \bibfield  {author} {\bibinfo {author} {\bibfnamefont {A.~K.}\ \bibnamefont
  {Boddeti}}, \bibinfo {author} {\bibfnamefont {J.}~\bibnamefont {Guan}},
  \bibinfo {author} {\bibfnamefont {T.}~\bibnamefont {Sentz}}, \bibinfo
  {author} {\bibfnamefont {X.}~\bibnamefont {Juarez}}, \bibinfo {author}
  {\bibfnamefont {W.}~\bibnamefont {Newman}}, \bibinfo {author} {\bibfnamefont
  {C.}~\bibnamefont {Cortes}}, \bibinfo {author} {\bibfnamefont {T.~W.}\
  \bibnamefont {Odom}}, \ and\ \bibinfo {author} {\bibfnamefont
  {Z.}~\bibnamefont {Jacob}},\ }\href {\doibase 10.1021/acs.nanolett.1c02835}
  {\bibfield  {journal} {\bibinfo  {journal} {Nano Letters}\ }\textbf {\bibinfo
  {volume} {22}},\ \bibinfo {pages} {22} (\bibinfo {year} {2022})}\BibitemShut
  {NoStop}%
\bibitem [{\citenamefont {Pennetta}\ \emph {et~al.}(2022)\citenamefont
  {Pennetta}, \citenamefont {Blaha}, \citenamefont {Johnson}, \citenamefont
  {Lechner}, \citenamefont {Schneeweiss}, \citenamefont {Volz},\ and\
  \citenamefont {Rauschenbeutel}}]{Pennetta2022}%
  \BibitemOpen
  \bibfield  {author} {\bibinfo {author} {\bibfnamefont {R.}~\bibnamefont
  {Pennetta}}, \bibinfo {author} {\bibfnamefont {M.}~\bibnamefont {Blaha}},
  \bibinfo {author} {\bibfnamefont {A.}~\bibnamefont {Johnson}}, \bibinfo
  {author} {\bibfnamefont {D.}~\bibnamefont {Lechner}}, \bibinfo {author}
  {\bibfnamefont {P.}~\bibnamefont {Schneeweiss}}, \bibinfo {author}
  {\bibfnamefont {J.}~\bibnamefont {Volz}}, \ and\ \bibinfo {author}
  {\bibfnamefont {A.}~\bibnamefont {Rauschenbeutel}},\ }\href {\doibase
  10.1103/PhysRevLett.128.073601} {\bibfield  {journal} {\bibinfo  {journal}
  {Phys. Rev. Lett.}\ }\textbf {\bibinfo {volume} {128}},\ \bibinfo {pages}
  {073601} (\bibinfo {year} {2022})}\BibitemShut {NoStop}%
\bibitem [{\citenamefont {Zanner}\ \emph {et~al.}(2022)\citenamefont {Zanner},
  \citenamefont {Orell}, \citenamefont {Schneider}, \citenamefont {Albert},
  \citenamefont {Oleschko}, \citenamefont {Juan}, \citenamefont {Silveri},\
  and\ \citenamefont {Kirchmair}}]{Zanner22}%
  \BibitemOpen
  \bibfield  {author} {\bibinfo {author} {\bibfnamefont {M.}~\bibnamefont
  {Zanner}}, \bibinfo {author} {\bibfnamefont {T.}~\bibnamefont {Orell}},
  \bibinfo {author} {\bibfnamefont {C.~M.~F.}\ \bibnamefont {Schneider}},
  \bibinfo {author} {\bibfnamefont {R.}~\bibnamefont {Albert}}, \bibinfo
  {author} {\bibfnamefont {S.}~\bibnamefont {Oleschko}}, \bibinfo {author}
  {\bibfnamefont {M.~L.}\ \bibnamefont {Juan}}, \bibinfo {author}
  {\bibfnamefont {M.}~\bibnamefont {Silveri}}, \ and\ \bibinfo {author}
  {\bibfnamefont {G.}~\bibnamefont {Kirchmair}},\ }\href
  {https://doi.org/10.1038/s41567-022-01527-w} {\bibfield  {journal} {\bibinfo
  {journal} {Nature Physics}\ }\textbf {\bibinfo {volume} {18}},\ \bibinfo
  {pages} {538} (\bibinfo {year} {2022})}\BibitemShut {NoStop}%
\bibitem [{\citenamefont {Solano}\ \emph
  {et~al.}(2017{\natexlab{b}})\citenamefont {Solano}, \citenamefont {Grover},
  \citenamefont {Hoffman}, \citenamefont {Ravets}, \citenamefont {Fatemi},
  \citenamefont {Orozco}, \citenamefont {Rolston},\ and\ \citenamefont
  {Optical~Nanofibers:}}]{PabloReview}%
  \BibitemOpen
  \bibfield  {author} {\bibinfo {author} {\bibfnamefont {P.}~\bibnamefont
  {Solano}}, \bibinfo {author} {\bibfnamefont {J.~A.}\ \bibnamefont {Grover}},
  \bibinfo {author} {\bibfnamefont {J.~E.}\ \bibnamefont {Hoffman}}, \bibinfo
  {author} {\bibfnamefont {S.}~\bibnamefont {Ravets}}, \bibinfo {author}
  {\bibfnamefont {F.~K.}\ \bibnamefont {Fatemi}}, \bibinfo {author}
  {\bibfnamefont {L.~A.}\ \bibnamefont {Orozco}}, \bibinfo {author}
  {\bibfnamefont {S.~L.}\ \bibnamefont {Rolston}}, \ and\ \bibinfo {author}
  {\bibfnamefont {A.}~\bibnamefont {Optical~Nanofibers:}},\ }\href {\doibase
  10.1016/bs.aamop.2017.02.003} {\bibfield  {journal} {\bibinfo  {journal}
  {Adv. At. Mol. Opt. Phys.}\ }\textbf {\bibinfo {volume} {66}},\ \bibinfo
  {pages} {439} (\bibinfo {year} {2017}{\natexlab{b}})}\BibitemShut {NoStop}%
\bibitem [{\citenamefont {Chang}\ \emph {et~al.}(2012)\citenamefont {Chang},
  \citenamefont {Jiang}, \citenamefont {Gorshkov},\ and\ \citenamefont
  {Kimble}}]{ChangNJP}%
  \BibitemOpen
  \bibfield  {author} {\bibinfo {author} {\bibfnamefont {D.~E.}\ \bibnamefont
  {Chang}}, \bibinfo {author} {\bibfnamefont {L.}~\bibnamefont {Jiang}},
  \bibinfo {author} {\bibfnamefont {A.~V.}\ \bibnamefont {Gorshkov}}, \ and\
  \bibinfo {author} {\bibfnamefont {H.~J.}\ \bibnamefont {Kimble}},\ }\href
  {\doibase 10.1088/1367-2630/14/6/063003} {\bibfield  {journal} {\bibinfo
  {journal} {New J. Phys.}\ }\textbf {\bibinfo {volume} {14}},\ \bibinfo
  {pages} {063003} (\bibinfo {year} {2012})}\BibitemShut {NoStop}%
\bibitem [{\citenamefont {Corzo}\ \emph {et~al.}(2016)\citenamefont {Corzo},
  \citenamefont {Gouraud}, \citenamefont {Chandra}, \citenamefont {Goban},
  \citenamefont {Sheremet}, \citenamefont {Kupriyanov},\ and\ \citenamefont
  {Laurat}}]{Corzo16}%
  \BibitemOpen
  \bibfield  {author} {\bibinfo {author} {\bibfnamefont {N.~V.}\ \bibnamefont
  {Corzo}}, \bibinfo {author} {\bibfnamefont {B.}~\bibnamefont {Gouraud}},
  \bibinfo {author} {\bibfnamefont {A.}~\bibnamefont {Chandra}}, \bibinfo
  {author} {\bibfnamefont {A.}~\bibnamefont {Goban}}, \bibinfo {author}
  {\bibfnamefont {A.~S.}\ \bibnamefont {Sheremet}}, \bibinfo {author}
  {\bibfnamefont {D.~V.}\ \bibnamefont {Kupriyanov}}, \ and\ \bibinfo {author}
  {\bibfnamefont {J.}~\bibnamefont {Laurat}},\ }\href {\doibase
  10.1103/PhysRevLett.117.133603} {\bibfield  {journal} {\bibinfo  {journal}
  {Phys. Rev. Lett.}\ }\textbf {\bibinfo {volume} {117}},\ \bibinfo {pages}
  {133603} (\bibinfo {year} {2016})}\BibitemShut {NoStop}%
\bibitem [{\citenamefont {Mirhosseini}\ \emph {et~al.}(2019)\citenamefont
  {Mirhosseini}, \citenamefont {Kim}, \citenamefont {Zhang}, \citenamefont
  {Sipahigil}, \citenamefont {Dieterle}, \citenamefont {Keller}, \citenamefont
  {Asenjo-Garcia}, \citenamefont {Chang},\ and\ \citenamefont
  {Painter}}]{Mirhosseini19}%
  \BibitemOpen
  \bibfield  {author} {\bibinfo {author} {\bibfnamefont {M.}~\bibnamefont
  {Mirhosseini}}, \bibinfo {author} {\bibfnamefont {E.}~\bibnamefont {Kim}},
  \bibinfo {author} {\bibfnamefont {X.}~\bibnamefont {Zhang}}, \bibinfo
  {author} {\bibfnamefont {A.}~\bibnamefont {Sipahigil}}, \bibinfo {author}
  {\bibfnamefont {P.~B.}\ \bibnamefont {Dieterle}}, \bibinfo {author}
  {\bibfnamefont {A.~J.}\ \bibnamefont {Keller}}, \bibinfo {author}
  {\bibfnamefont {A.}~\bibnamefont {Asenjo-Garcia}}, \bibinfo {author}
  {\bibfnamefont {D.~E.}\ \bibnamefont {Chang}}, \ and\ \bibinfo {author}
  {\bibfnamefont {O.}~\bibnamefont {Painter}},\ }\href {\doibase
  10.1038/s41586-019-1196-1} {\bibfield  {journal} {\bibinfo  {journal}
  {Nature}\ }\textbf {\bibinfo {volume} {569}},\ \bibinfo {pages} {692}
  (\bibinfo {year} {2019})}\BibitemShut {NoStop}%
\bibitem [{\citenamefont {Ostermann}\ \emph {et~al.}(2013)\citenamefont
  {Ostermann}, \citenamefont {Ritsch},\ and\ \citenamefont
  {Genes}}]{Ostermann13}%
  \BibitemOpen
  \bibfield  {author} {\bibinfo {author} {\bibfnamefont {L.}~\bibnamefont
  {Ostermann}}, \bibinfo {author} {\bibfnamefont {H.}~\bibnamefont {Ritsch}}, \
  and\ \bibinfo {author} {\bibfnamefont {C.}~\bibnamefont {Genes}},\ }\href
  {\doibase 10.1103/PhysRevLett.111.123601} {\bibfield  {journal} {\bibinfo
  {journal} {Phys. Rev. Lett.}\ }\textbf {\bibinfo {volume} {111}},\ \bibinfo
  {pages} {123601} (\bibinfo {year} {2013})}\BibitemShut {NoStop}%
\bibitem [{\citenamefont {Henriet}\ \emph {et~al.}(2019)\citenamefont
  {Henriet}, \citenamefont {Douglas}, \citenamefont {Chang},\ and\
  \citenamefont {Albrecht}}]{Henriet19}%
  \BibitemOpen
  \bibfield  {author} {\bibinfo {author} {\bibfnamefont {L.}~\bibnamefont
  {Henriet}}, \bibinfo {author} {\bibfnamefont {J.~S.}\ \bibnamefont
  {Douglas}}, \bibinfo {author} {\bibfnamefont {D.~E.}\ \bibnamefont {Chang}},
  \ and\ \bibinfo {author} {\bibfnamefont {A.}~\bibnamefont {Albrecht}},\
  }\href {\doibase 10.1103/PhysRevA.99.023802} {\bibfield  {journal} {\bibinfo
  {journal} {Phys. Rev. A}\ }\textbf {\bibinfo {volume} {99}},\ \bibinfo
  {pages} {023802} (\bibinfo {year} {2019})}\BibitemShut {NoStop}%
\bibitem [{\citenamefont {Manzoni}\ \emph {et~al.}(2018)\citenamefont
  {Manzoni}, \citenamefont {Moreno-Cardoner}, \citenamefont {Asenjo-Garcia},
  \citenamefont {Porto}, \citenamefont {Gorshkov},\ and\ \citenamefont
  {Chang}}]{Manzoni_2018}%
  \BibitemOpen
  \bibfield  {author} {\bibinfo {author} {\bibfnamefont {M.~T.}\ \bibnamefont
  {Manzoni}}, \bibinfo {author} {\bibfnamefont {M.}~\bibnamefont
  {Moreno-Cardoner}}, \bibinfo {author} {\bibfnamefont {A.}~\bibnamefont
  {Asenjo-Garcia}}, \bibinfo {author} {\bibfnamefont {J.~V.}\ \bibnamefont
  {Porto}}, \bibinfo {author} {\bibfnamefont {A.~V.}\ \bibnamefont {Gorshkov}},
  \ and\ \bibinfo {author} {\bibfnamefont {D.~E.}\ \bibnamefont {Chang}},\
  }\href {https://doi.org/10.1088/1367-2630/aadb74} {\bibfield  {journal}
  {\bibinfo  {journal} {New Journal of Physics}\ }\textbf {\bibinfo {volume}
  {20}},\ \bibinfo {pages} {083048} (\bibinfo {year} {2018})}\BibitemShut
  {NoStop}%
\bibitem [{\citenamefont {Din\c{c}}\ and\ \citenamefont
  {Bra\'{n}czyk}(2019)}]{Dinc19}%
  \BibitemOpen
  \bibfield  {author} {\bibinfo {author} {\bibfnamefont {F.}~\bibnamefont
  {Din\c{c}}}\ and\ \bibinfo {author} {\bibfnamefont {A.~M.}\ \bibnamefont
  {Bra\'{n}czyk}},\ }\href {\doibase 10.1103/PhysRevResearch.1.032042}
  {\bibfield  {journal} {\bibinfo  {journal} {Phys. Rev. Research}\ }\textbf
  {\bibinfo {volume} {1}},\ \bibinfo {pages} {032042(R)} (\bibinfo {year}
  {2019})}\BibitemShut {NoStop}%
\bibitem [{\citenamefont {Rist}\ \emph {et~al.}(2008)\citenamefont {Rist},
  \citenamefont {Eschner}, \citenamefont {Hennrich},\ and\ \citenamefont
  {Morigi}}]{Rist08}%
  \BibitemOpen
  \bibfield  {author} {\bibinfo {author} {\bibfnamefont {S.}~\bibnamefont
  {Rist}}, \bibinfo {author} {\bibfnamefont {J.}~\bibnamefont {Eschner}},
  \bibinfo {author} {\bibfnamefont {M.}~\bibnamefont {Hennrich}}, \ and\
  \bibinfo {author} {\bibfnamefont {G.}~\bibnamefont {Morigi}},\ }\href
  {\doibase 10.1103/PhysRevA.78.013808} {\bibfield  {journal} {\bibinfo
  {journal} {Phys. Rev. A}\ }\textbf {\bibinfo {volume} {78}},\ \bibinfo
  {pages} {013808} (\bibinfo {year} {2008})}\BibitemShut {NoStop}%
\bibitem [{\citenamefont {Sinha}\ \emph
  {et~al.}(2020{\natexlab{a}})\citenamefont {Sinha}, \citenamefont
  {Gonz\'alez-Tudela}, \citenamefont {Lu},\ and\ \citenamefont
  {Solano}}]{Sinha20}%
  \BibitemOpen
  \bibfield  {author} {\bibinfo {author} {\bibfnamefont {K.}~\bibnamefont
  {Sinha}}, \bibinfo {author} {\bibfnamefont {A.}~\bibnamefont
  {Gonz\'alez-Tudela}}, \bibinfo {author} {\bibfnamefont {Y.}~\bibnamefont
  {Lu}}, \ and\ \bibinfo {author} {\bibfnamefont {P.}~\bibnamefont {Solano}},\
  }\href {\doibase 10.1103/PhysRevA.102.043718} {\bibfield  {journal} {\bibinfo
   {journal} {Phys. Rev. A}\ }\textbf {\bibinfo {volume} {102}},\ \bibinfo
  {pages} {043718} (\bibinfo {year} {2020}{\natexlab{a}})}\BibitemShut
  {NoStop}%
\bibitem [{\citenamefont {Sinha}\ \emph
  {et~al.}(2020{\natexlab{b}})\citenamefont {Sinha}, \citenamefont {Meystre},
  \citenamefont {Goldschmidt}, \citenamefont {Fatemi}, \citenamefont
  {Rolston},\ and\ \citenamefont {Solano}}]{Sinha_2020}%
  \BibitemOpen
  \bibfield  {author} {\bibinfo {author} {\bibfnamefont {K.}~\bibnamefont
  {Sinha}}, \bibinfo {author} {\bibfnamefont {P.}~\bibnamefont {Meystre}},
  \bibinfo {author} {\bibfnamefont {E.~A.}\ \bibnamefont {Goldschmidt}},
  \bibinfo {author} {\bibfnamefont {F.~K.}\ \bibnamefont {Fatemi}}, \bibinfo
  {author} {\bibfnamefont {S.~L.}\ \bibnamefont {Rolston}}, \ and\ \bibinfo
  {author} {\bibfnamefont {P.}~\bibnamefont {Solano}},\ }\href {\doibase
  10.1103/PhysRevLett.124.043603} {\bibfield  {journal} {\bibinfo  {journal}
  {Phys. Rev. Lett.}\ }\textbf {\bibinfo {volume} {124}},\ \bibinfo {pages}
  {043603} (\bibinfo {year} {2020}{\natexlab{b}})}\BibitemShut {NoStop}%
\bibitem [{\citenamefont {Sinha}\ \emph {et~al.}(2019)\citenamefont {Sinha},
  \citenamefont {Meystre},\ and\ \citenamefont {Solano}}]{Sinha2019}%
  \BibitemOpen
  \bibfield  {author} {\bibinfo {author} {\bibfnamefont {K.}~\bibnamefont
  {Sinha}}, \bibinfo {author} {\bibfnamefont {P.}~\bibnamefont {Meystre}}, \
  and\ \bibinfo {author} {\bibfnamefont {P.}~\bibnamefont {Solano}},\ }\href
  {\doibase 10.1117/12.2530927} {\bibfield  {journal} {\bibinfo  {journal}
  {Nanophotonic Materials, Devices, and Systems}\ }\textbf {\bibinfo {volume}
  {11091}},\ \bibinfo {pages} {53 } (\bibinfo {year} {2019})}\BibitemShut
  {NoStop}%
\bibitem [{\citenamefont {Calaj\'o}\ \emph {et~al.}(2019)\citenamefont
  {Calaj\'o}, \citenamefont {Fang}, \citenamefont {Baranger},\ and\
  \citenamefont {Ciccarello}}]{Calajo_2019}%
  \BibitemOpen
  \bibfield  {author} {\bibinfo {author} {\bibfnamefont {G.}~\bibnamefont
  {Calaj\'o}}, \bibinfo {author} {\bibfnamefont {Y.-L.~L.}\ \bibnamefont
  {Fang}}, \bibinfo {author} {\bibfnamefont {H.~U.}\ \bibnamefont {Baranger}},
  \ and\ \bibinfo {author} {\bibfnamefont {F.}~\bibnamefont {Ciccarello}},\
  }\href {\doibase 10.1103/PhysRevLett.122.073601} {\bibfield  {journal}
  {\bibinfo  {journal} {Phys. Rev. Lett.}\ }\textbf {\bibinfo {volume} {122}},\
  \bibinfo {pages} {073601} (\bibinfo {year} {2019})}\BibitemShut {NoStop}%
\bibitem [{\citenamefont {Trivedi}\ \emph {et~al.}(2021)\citenamefont
  {Trivedi}, \citenamefont {Malz}, \citenamefont {Sun}, \citenamefont {Fan},\
  and\ \citenamefont {Vu\ifmmode \check{c}\else
  \v{c}\fi{}kovi\ifmmode~\acute{c}\else \'{c}\fi{}}}]{Trivedi21}%
  \BibitemOpen
  \bibfield  {author} {\bibinfo {author} {\bibfnamefont {R.}~\bibnamefont
  {Trivedi}}, \bibinfo {author} {\bibfnamefont {D.}~\bibnamefont {Malz}},
  \bibinfo {author} {\bibfnamefont {S.}~\bibnamefont {Sun}}, \bibinfo {author}
  {\bibfnamefont {S.}~\bibnamefont {Fan}}, \ and\ \bibinfo {author}
  {\bibfnamefont {J.}~\bibnamefont {Vu\ifmmode \check{c}\else
  \v{c}\fi{}kovi\ifmmode~\acute{c}\else \'{c}\fi{}}},\ }\href {\doibase
  10.1103/PhysRevA.104.013705} {\bibfield  {journal} {\bibinfo  {journal}
  {Phys. Rev. A}\ }\textbf {\bibinfo {volume} {104}},\ \bibinfo {pages}
  {013705} (\bibinfo {year} {2021})}\BibitemShut {NoStop}%
\bibitem [{\citenamefont {Yao}\ and\ \citenamefont {Hughes}(2009)}]{Hughes09}%
  \BibitemOpen
  \bibfield  {author} {\bibinfo {author} {\bibfnamefont {P.}~\bibnamefont
  {Yao}}\ and\ \bibinfo {author} {\bibfnamefont {S.}~\bibnamefont {Hughes}},\
  }\href {\doibase 10.1364/OE.17.011505} {\bibfield  {journal} {\bibinfo
  {journal} {Opt. Express}\ }\textbf {\bibinfo {volume} {17}},\ \bibinfo
  {pages} {11505} (\bibinfo {year} {2009})}\BibitemShut {NoStop}%
\bibitem [{\citenamefont {Hegerfeldt}\ and\ \citenamefont
  {Plenio}(1994)}]{Hegerfeldt_1994}%
  \BibitemOpen
  \bibfield  {author} {\bibinfo {author} {\bibfnamefont {G.~C.}\ \bibnamefont
  {Hegerfeldt}}\ and\ \bibinfo {author} {\bibfnamefont {M.~B.}\ \bibnamefont
  {Plenio}},\ }\href {\doibase 10.1088/0954-8998/6/1/003} {\bibfield  {journal}
  {\bibinfo  {journal} {Quantum Opt.}\ }\textbf {\bibinfo {volume} {6}},\
  \bibinfo {pages} {15} (\bibinfo {year} {1994})}\BibitemShut {NoStop}%
\bibitem [{\citenamefont {Srinivasan}\ \emph {et~al.}(2011)\citenamefont
  {Srinivasan}, \citenamefont {Hoffman}, \citenamefont {Gambetta},\ and\
  \citenamefont {Houck}}]{Srinivasan11}%
  \BibitemOpen
  \bibfield  {author} {\bibinfo {author} {\bibfnamefont {S.~J.}\ \bibnamefont
  {Srinivasan}}, \bibinfo {author} {\bibfnamefont {A.~J.}\ \bibnamefont
  {Hoffman}}, \bibinfo {author} {\bibfnamefont {J.~M.}\ \bibnamefont
  {Gambetta}}, \ and\ \bibinfo {author} {\bibfnamefont {A.~A.}\ \bibnamefont
  {Houck}},\ }\href {\doibase 10.1103/PhysRevLett.106.083601} {\bibfield
  {journal} {\bibinfo  {journal} {Phys. Rev. Lett.}\ }\textbf {\bibinfo
  {volume} {106}},\ \bibinfo {pages} {083601} (\bibinfo {year}
  {2011})}\BibitemShut {NoStop}%
\bibitem [{\citenamefont {Din\c{c}}\ \emph {et~al.}(2019)\citenamefont
  {Din\c{c}}, \citenamefont {Bra\'{n}czyk},\ and\ \citenamefont
  {Ercan}}]{Dinc18}%
  \BibitemOpen
  \bibfield  {author} {\bibinfo {author} {\bibfnamefont {F.}~\bibnamefont
  {Din\c{c}}}, \bibinfo {author} {\bibfnamefont {A.~M.}\ \bibnamefont
  {Bra\'{n}czyk}}, \ and\ \bibinfo {author} {\bibfnamefont {I.}~\bibnamefont
  {Ercan}},\ }\href {https://doi.org/10.22331/q-2019-12-09-213} {\bibfield
  {journal} {\bibinfo  {journal} {Quantum}\ }\textbf {\bibinfo {volume} {3}},\
  \bibinfo {pages} {213} (\bibinfo {year} {2019})}\BibitemShut {NoStop}%
\bibitem [{\citenamefont {Asenjo-Garcia}\ \emph {et~al.}(2019)\citenamefont
  {Asenjo-Garcia}, \citenamefont {Kimble},\ and\ \citenamefont
  {Chang}}]{Asenjo19}%
  \BibitemOpen
  \bibfield  {author} {\bibinfo {author} {\bibfnamefont {A.}~\bibnamefont
  {Asenjo-Garcia}}, \bibinfo {author} {\bibfnamefont {H.~J.}\ \bibnamefont
  {Kimble}}, \ and\ \bibinfo {author} {\bibfnamefont {D.~E.}\ \bibnamefont
  {Chang}},\ }\href {https://doi.org/10.1073/pnas.1911467116} {\bibfield
  {journal} {\bibinfo  {journal} {Proceedings of the National Academy of
  Sciences}\ }\textbf {\bibinfo {volume} {116}},\ \bibinfo {pages} {25503}
  (\bibinfo {year} {2019})}\BibitemShut {NoStop}%
\bibitem [{\citenamefont {Cerf}\ \emph {et~al.}(2002)\citenamefont {Cerf},
  \citenamefont {Bourennane}, \citenamefont {Karlsson},\ and\ \citenamefont
  {Gisin}}]{Cerf02}%
  \BibitemOpen
  \bibfield  {author} {\bibinfo {author} {\bibfnamefont {N.~J.}\ \bibnamefont
  {Cerf}}, \bibinfo {author} {\bibfnamefont {M.}~\bibnamefont {Bourennane}},
  \bibinfo {author} {\bibfnamefont {A.}~\bibnamefont {Karlsson}}, \ and\
  \bibinfo {author} {\bibfnamefont {N.}~\bibnamefont {Gisin}},\ }\href
  {\doibase 10.1103/PhysRevLett.88.127902} {\bibfield  {journal} {\bibinfo
  {journal} {Phys. Rev. Lett.}\ }\textbf {\bibinfo {volume} {88}},\ \bibinfo
  {pages} {127902} (\bibinfo {year} {2002})}\BibitemShut {NoStop}%
\bibitem [{\citenamefont {Vaziri}\ \emph {et~al.}(2002)\citenamefont {Vaziri},
  \citenamefont {Weihs},\ and\ \citenamefont {Zeilinger}}]{Vaziri02}%
  \BibitemOpen
  \bibfield  {author} {\bibinfo {author} {\bibfnamefont {A.}~\bibnamefont
  {Vaziri}}, \bibinfo {author} {\bibfnamefont {G.}~\bibnamefont {Weihs}}, \
  and\ \bibinfo {author} {\bibfnamefont {A.}~\bibnamefont {Zeilinger}},\ }\href
  {\doibase 10.1103/PhysRevLett.89.240401} {\bibfield  {journal} {\bibinfo
  {journal} {Phys. Rev. Lett.}\ }\textbf {\bibinfo {volume} {89}},\ \bibinfo
  {pages} {240401} (\bibinfo {year} {2002})}\BibitemShut {NoStop}%
\bibitem [{\citenamefont {Bechmann-Pasquinucci}\ and\ \citenamefont
  {Tittel}(2000)}]{Tittel2000}%
  \BibitemOpen
  \bibfield  {author} {\bibinfo {author} {\bibfnamefont {H.}~\bibnamefont
  {Bechmann-Pasquinucci}}\ and\ \bibinfo {author} {\bibfnamefont
  {W.}~\bibnamefont {Tittel}},\ }\href {\doibase 10.1103/PhysRevA.61.062308}
  {\bibfield  {journal} {\bibinfo  {journal} {Phys. Rev. A}\ }\textbf {\bibinfo
  {volume} {61}},\ \bibinfo {pages} {062308} (\bibinfo {year}
  {2000})}\BibitemShut {NoStop}%
\end{thebibliography}%

\end{document}